\documentclass[twocolumn,longbib]{aastex701}

\usepackage{lipsum}
\usepackage{hyperref}
\usepackage{multirow}
\usepackage{xfrac}
\usepackage{CJK}

\newcommand{\update}[1]{\textcolor{black}{#1}}
\newcommand{\updatetwo}[1]{\textcolor{black}{#1}}

\makeatletter
\newcommand\footnoteref[1]{\protected@xdef\@thefnmark{\ref{#1}}\@footnotemark}
\makeatother

\graphicspath{{./}{figures/}{103p/}{chiron/}}

\begin{document}
\begin{CJK*}{UTF8}{gbsn}

\title{Multi-year Ground-Based Survey Photometry of Active Comet 103P/Hartley 2 and Centaur (2060) Chiron: A Tale of Two Comets in the Pre-LSST Era}

\author[0000-0001-9505-1131]{Joseph Murtagh}\altaffiliation{DiRAC Postdoctoral Fellow}
\affiliation{Astrophysics Research Centre, School of Mathematics and Physics, Queen's University Belfast, BT7 1NN, UK}
\affiliation{Department of Astronomy and the DIRAC Institute, University of Washington, 3910 15th Avenue NE, Seattle, WA 98195, USA}
\email[show]{murtagh@uw.edu}
\correspondingauthor{Joseph Murtagh}

\author[0000-0003-4365-1455]{Megan E. Schwamb}
\affiliation{Astrophysics Research Centre, School of Mathematics and Physics, Queen's University Belfast, BT7 1NN, UK}
\email{m.schwamb@qub.ac.uk}

\author[0000-0002-6702-7676]{Michael S. P. Kelley}
\affiliation{Department of Astronomy, University of Maryland, College Park, MD 20742-0001, USA}
\email{msk@astro.umd.edu}

\author[0000-0003-2781-6897]{Matthew M. Knight}
\affiliation{Volgenau Department of Physics, United States Naval Academy, Annapolis, MD 21402, USA}
\email{knight@usna.edu}

\author[0009-0004-3836-8166]{Thomas Lehmann}
\affiliation{School Observatory Friedrich-Schiller-Gymnasium, Weimar, Germany}
\email{t.lehmann@mailbox.org}

\author[0000-0002-6158-6370]{Giannantonio Milani}
\affiliation{CARA Project - INAF - Osservatorio Astronomico di Trieste, Via Tiepolo 11, I-34143 Trieste, Italy}
\email{giannantonio.milani@gmail.com}

\author[0000-0003-4094-9408]{A. Fraser Gillan}
\affiliation{Astrophysics Division, National Centre for Nuclear Research, Pasteura 7, 02-093 Warsaw, Poland}
\email{andrew.gillan@ncbj.gov.pl}

\author[0000-0003-0250-9911]{Alan Fitzsimmons}
\affiliation{Astrophysics Research Centre, School of Mathematics and Physics, Queen's University Belfast, BT7 1NN, UK}
\email{a.fitzsimmons@qub.ac.uk}

\author[0009-0007-8602-2954]{Ryan R. Lyttle}
\affiliation{Astrophysics Research Centre, School of Mathematics and Physics, Queen's University Belfast, BT7 1NN, UK}
\email{rlyttle09@qub.ac.uk}

\author[0000-0002-1105-7980]{Matthew M. Dobson}
\affiliation{Astrophysics Research Centre, School of Mathematics and Physics, Queen's University Belfast, BT7 1NN, UK}
\email{mdobson04@qub.ac.uk}

\author[0000-0002-4043-6445]{Carrie E. Holt}
\affiliation{Las Cumbres Observatory, 6740 Cortona Drive, Suite 102, Goleta, CA 93117, USA}
\email{cholt@lco.global}

\author[0000-0002-3818-7769]{Tim Lister}
\affiliation{Las Cumbres Observatory, 6740 Cortona Drive, Suite 102, Goleta, CA 93117, USA}
\email{tlister@lco.global}

\author[0000-0002-1278-5998]{Joseph P. Chatelain}
\affiliation{Las Cumbres Observatory, 6740 Cortona Drive, Suite 102, Goleta, CA 93117, USA}
\email{jchatelain@lco.global}

\author[0000-0002-4439-1539]{Sarah Greenstreet}
\affiliation{Rubin Observatory/NSF NOIRLab, 950 North Cherry Avenue, Tucson, AZ 85719, USA}
\affiliation{Department of Astronomy and the DIRAC Institute, University of Washington, 3910 15th Avenue NE, Seattle, WA 98195, USA}
\email{sarahjg@uw.edu}

\author[0000-0002-4838-7676]{Quanzhi Ye (叶泉志)}
\affiliation{Department of Astronomy, University of Maryland, College Park, MD 20742-0001, USA}
\affiliation{Center for Space Physics, Boston University, 725 Commonwealth Avenue, Boston, MA 02215, USA}
\email{qye@umd.edu}

\author[0000-0001-7225-9271]{Henry H. Hsieh}
\affiliation{Planetary Science Institute, 1700 East Fort Lowell Road, Suite 106, Tucson, AZ 85719, USA}
\email{hhsieh@psi.edu}

\author[0000-0002-2668-7248]{Denis Bodewits}
\affiliation{Physics Department, Edmund C. Leach Science Center, Auburn University, Auburn, AL 36832, USA}
\email{dennis@auburn.edu}


\author[0000-0001-9152-6224]{Tracy X. Chen}
\affiliation{IPAC, California Institute of Technology, 1200 E. California Blvd, Pasadena, CA 91125, USA}
\email{xchen@ipac.caltech.edu}

\author[0000-0002-5884-7867]{Richard G. Dekany}
\affiliation{Caltech Optical Observatories, California Institute of Technology, Pasadena, CA  91125, USA}
\email{rgd@astro.caltech.edu}

\author[0000-0002-3168-0139]{Matthew J. Graham}
\affiliation{California Institute of Technology, 1200 E. California Blvd, Pasadena, CA 91125, USA}
\email{mjg@caltech.edu}

\author[0000-0001-5668-3507]{Steven L. Groom}
\affiliation{IPAC, California Institute of Technology, 1200 E. California Blvd, Pasadena, CA 91125, USA}
\email{sgroom@caltech.edu}

\author[0000-0003-3367-3415]{George Helou}
\affiliation{IPAC, California Institute of Technology, 1200 E. California Blvd, Pasadena, CA 91125, USA}
\email{gxh@ipac.caltech.edu}

\author[0000-0002-5619-4938]{Mansi M. Kasliwal}
\affiliation{Division of Physics, Mathematics, and Astronomy, California Institute of Technology, Pasadena, CA 91125, USA}
\email{mansi@astro.caltech.edu}

\author[0000-0003-2242-0244]{Ashish~A.~Mahabal}
\affiliation{Division of Physics, Mathematics, and Astronomy, California Institute of Technology, Pasadena, CA 91125, USA}
\affiliation{Center for Data Driven Discovery, California Institute of Technology, Pasadena, CA 91125, USA}
\email{aam@astro.caltech.edu}

\author[0000-0002-0387-370X]{Reed Riddle}
\affiliation{Caltech Optical Observatories, California Institute of Technology, Pasadena, CA 91125, USA}
\email{riddle@caltech.edu}

\begin{abstract}

Comets and Centaurs trace the evolution of trans-Neptunian \updatetwo{objects (TNOs)} into the inner solar system. Their activity reflects the interplay between volatile sublimation, dust dynamics, and, in some cases, ring scattering. Yet the long-term behavior of individual objects remains less constrained. \updatetwo{To probe this evolutionary transition, we} use wide-field survey photometry from the Asteroid Terrestrial-impact Last Alert System, Zwicky Transient Facility, and targeted Las Cumbres Observatory observations \updatetwo{of} the Jupiter-family comet \update{(JFC)} 103P/Hartley 2 during its 2023/24 apparition, and the Centaur (2060) Chiron across 2020-2025, including its 2021 outburst. For 103P, heliocentric activity slopes are asymmetric about perihelion, with a steep inbound index ($n_{r,\rm pre}=-3.48\pm0.08$) and flatter outbound value ($n_{r,\rm post}=-1.16\pm0.04$), consistent with enhanced relative dust contribution post-perihelion. \updatetwo{With} reduced brightness \updatetwo{versus prior} apparitions, this behavior matches reported secular fading trends. \updatetwo{Dust} mass-loss rates are $\sim4$-16~kg~s$^{-1}$ \updatetwo{for} assumed grain properties. Colors exhibit a blueward trend near perihelion, consistent with enhanced gas contamination of the $g$-band, \updatetwo{with possible} phase-dependent scattering. Periodogram analysis recovers a $\sim18.7$~hr activity-linked period near perihelion. For Chiron, subtracting a quiescent baseline reveals exponential decay from the 2021 outburst \updatetwo{on} a $\sim1.4$~yr \updatetwo{timescale}. Seasonal phase curves flatten from $\beta_o=0.150\pm0.034$~mag~deg$^{-1}$ in 2021 to $\lesssim0.09$~mag~deg$^{-1}$ by 2023-2025, converging with quiescent behavior. Broad-band colors remain unchanged at ATLAS ($c-o$) = 0.22$\pm$0.09~mag. The extended duration of this activity suggests a new epoch of persistent, low-level activity and/or evolving ring-scattering. \updatetwo{These two objects bracket the TNO-to-JFC evolutionary sequence, with 103P near the volatile-depleted end, and Chiron still volatile-rich and capable of episodic activity.}

\end{abstract}



\section{Introduction} 

Small bodies in the outer solar system serve as nearly pristine remnants of the early stages of planetary formation, preserving critical information about the conditions and processes that shaped the solar system over 4.6 billion years. Amongst these objects, Jupiter-family comets (JFCs) and Centaurs represent particularly important populations that bridge our understanding between the distant trans-Neptunian region and the inner solar system. Most JFCs are thought to originate from the scattering disk as part of the trans-Neptunian region, whereby gravitational interactions with Neptune inject them on to unstable, inwards evolving orbits \citep{holman93, duncan97, levison97, volk08, jewitt09}. These objects can then spend millions of years as Centaurs between Jupiter and Neptune \citep{tiscareno03, disisto07}, with most eventually ejected from the solar system by further giant planet interactions, and only a small fraction transitioning into active JFCS \citep{horner04}. 

Once closer to the Sun, JFCs begin to display their characteristic cometary activity, dominated by water ice sublimation \citep{meech04, womack17}. Whilst Centaurs also display cometary activity, their large heliocentric distances mean that there must be some (as-of-yet-unknown) alternative driver for activity than water ice sublimation — recent James Webb Space Telescope (JWST) observations have detected CO$_2$ gas emission in the comae of active Centaurs \citep{harringtonpinto23, pinillaalonso24}, reinforcing this idea. Numerical studies hint that activity may be a result of sudden drops in perihelia distances inducing exothermic phase changes in the structure of internal ices from amorphous to crystalline \citep{fernandez18, lilly21, lilly24}. Throughout both of their dynamical evolutions, their surfaces are modified by solar radiation, thermal cycling, out-gassing, and resurfacing events, leading to changes in albedo, surface color, and rotational properties \citep[e.g.][]{prialnik04, jewitt09, melita12, peixinho12, kokotanekova17}. Observations of the photometric behavior of both Centaurs and JFCs via light curves, phase functions, and color evolution across long temporal baselines can therefore help directly probe the processes that transform pristine trans-Neptunian material into the short-period comets observed today.

Comet 103P/Hartley 2 (hereafter, 103P) and Centaur (2060) Chiron (hereafter, Chiron) represent two cases of these classes that can offer us just such unique insights in the evolution and activity mechanisms of primitive objects. 103P is a JFC on a 6.48~year orbit\footnote{As per the \href{https://minorplanetcenter.net/db_search/show_object?utf8=\%E2\%9C\%93&object_id=103P}{Minor Planet Center record for 103P/Hartley 2}}, with a history of repeated near-perihelion activity since its discovery in 1986 \citep{hartley86}. It has been observed extensively through ground based observations \citep[e.g.][]{lowry01, snodgrass08, mazzotta08, meech11, knight13, mueller13}, and from space via both satellite \citep[][]{jorda00, groussin04, bauer11, combi11} and \textit{in situ} observations with Extrasolar Planet Observation and Deep Impact Extended Investigation \citep[EPOXI;][]{ahearn11}. The EPOXI flyby in 2010 revealed 103P's elongated, bi-lobed nucleus shape \citep[long axis $\sim2.33$~km, short axis $\sim0.69$~km,][]{groussin13, thomas13}, evidence for non-principal-axis rotation \citep[NPA or ``tumbling'',][]{ahearn11, knight11, samarasinha11, belton13b}, and asymmetric volatile driven jets \citep{belton13, brucksyal13, steckloff16, belton17, vaughan17}. Subsequent monitoring revealed evidence for a lengthening of the apparent rotational period from $\sim16$~hr to $\sim19$~hr over the course of the 2010 apparition \citep{knight15}, with more recent observations suggesting the spin-down rate may have slowed significantly by the 2023/24 return \citep{lehmann25}. EPOXI also identified large, $\gtrsim$cm sized icy grains within the coma \citep{ahearn11, hermalyn13, kelley13, kelley15}, that drive water production rates beyond what the nucleus surface area alone can sustain, earning 103P its `hyperactive' classification\citep{groussin04}. More recently, \cite{graykowski25} identified a gradual fading of $\sim41\%$ per apparition across all observable returns since 1991, suggesting 103P may have lost its hyperactive status by 2017. 

In contrast, Chiron, a Centaur on a $\sim50$~year orbit between Saturn and Uranus\footnote{As per the \href{https://minorplanetcenter.net/db_search/show_object?utf8=\%E2\%9C\%93&object_id=2060}{Minor Planet Center record for (2060) Chiron}}, displays low-amplitude rotational variations \citep[$\lesssim0.1$ mag, ][]{groussin04b, ortiz15, sickafoose20, bragaribas23} consistent with a Jacobi ellipsoid with volume-equivalent radius $R_{\mathrm{vol}}=98$~km \citep{bragaribas23}. Chiron is known for episodic historical photometric variability, indicative of outbursts \citep{meech89, bus91, bus01, dobson21}, cometary activity \citep{tholen88, luu90, bus96, lazzaro97, duffard02, dobson23, dobson24, pinillaalonso24, wong26}, and the presence of a ring system \citep{ortiz15, pereira25}, which recent observations suggest is still actively forming and evolving. Recently, Chiron experienced a significant brightening in 2021, increasing by $\sim1$ mag over a short period of time and lasting for its entire 2021-2022 observing season \citep{dobson21, betzler23, ortiz23, dobson24}. This event was analyzed in detail by \cite{dobson24} who ruled out two possible explanations: (1) increased scattering from the ring system, since the observed strength and short duration of the brightening do not match model predictions, and (2) surface heterogeneity, because Chiron's measured axial pole orientation and obliquity to Earth mean that the projected surface terrain visible during this time did not change enough to expose large bright regions. Whilst \cite{dobson24} found no visible extension of the continuum beyond the point-spread function (PSF), \update{JWST independently revealed distinct CH$_4$ and CO$_2$ gas coma morphologies on 2023 July 12 \citep{pinillaalonso24} and 2024 Jan 9 \citep{wong26},} confirming a new or enhanced epoch of cometary gas activity for Chiron.

Long observational baselines are critical in understanding the processes driving activity in both comets and Centaurs, which can operate across a range of timescales from short-lived outbursts \citep[e.g.][]{belton13c, ishiguro16} to longer-term evolution of rotational spin-state \citep[e.g.][]{samarasinha04, kokotanekova17, donaldson23}, surface properties \citep[e.g.][]{bauer03, lamy04, peixinho12}, and volatile reservoirs \cite[e.g.][]{bockeleemorvan04, drahus17}. For 103P, repeated monitoring across successive perihelion passages can reveal whether its hyperactivity, rotation state, and coma morphology are stable over decades, or if they evolve measurably from orbit-to-orbit. In the case of Chiron, whose variability has historically been episodic and unpredictable, only extended coverage spanning many years can help constrain whether individual brightenings are isolated events or part of a recurring cycle of activity driven by internal or orbital factors \citep{jewitt09, lilly24}. Wide-field surveys now provide just such multi-year archives, with hundreds of photometric measurements per object enabling both epoch-to-epoch monitoring of JFCs and characterization of episodic Centaur brightenings. In particular, the Asteroid Terrestrial-impact Last Alert System \citep[ATLAS,][]{tonry18}, the Zwicky Transient Facility \citep[ZTF,][]{bellm19, graham19}, and complementary targeted follow-up from the Las Cumbres Observatory network \citep[LCO,][]{brown13}, accessed via the LCO Outbursting Objects Key Project \citep[LOOK,][]{lister22}, are well-suited to this purpose. Together, they provide the survey depth and targeted cadence necessary to trace both the gradual secular evolution of JFCs and the episodic activity of Centaurs over multi-year baselines.

In this work, we present a multi-year photometric analysis of Chiron and 103P's recent 2023/24 apparition using observations from ATLAS, ZTF, and LCO/LOOK, from 2020 to present day in order to probe their light curves, phase curves, surface color evolution, and activity and rotational periodicities in 103P specifically. In Section~\ref{sec:2} we present the details of the surveys used within this study, whilst Section~\ref{sec:3} explains the data reduction and processing steps for each telescope, and the calibrations applied to each dataset. Section~\ref{sec:4} presents the results for both objects together with their interpretations. \update{In Section~\ref{sec:5} we discuss the prospects for future wide-field surveys to perform similar work on active objects.} Finally, in Section~\ref{sec:6} we outline the conclusions of this study.

\section{Observations \& Data Reduction} \label{sec:2}

Here we briefly summarize the surveys (ATLAS, ZTF) with serendipitous imaging, and targeted observations (LCO/LOOK) of 103P and Chiron. Our observations of Chiron span six observing seasons from 2020 to 2025, covering the immediate pre-outburst, outburst, and post-outburst lightcurve, extending the dataset by two years beyond the study of \cite{dobson24} from 2024 to the end of the 2025/26 season\update{; throughout this window Chiron was near aphelion ($r_\mathrm{H}\approx18.87$~au, reached 2021 May), with $r_\mathrm{H}$ ranging from $\approx18.85$~au at the start of the dataset to $\approx18.4$~au by 2026 January}. Our 103P observations cover the most recent apparition from May 2023 to May 2024\update{; 103P reached perihelion on 2023 October 12 at $r_\mathrm{H}\approx1.06$~au, with observations spanning $r_\mathrm{H} \approx2.00$~au inbound (2023 May 28) to $r_\mathrm{H}\approx2.8$~au outbound (2024 May 27)}, in line with the recent study from \cite{lehmann25}. The summary of the number of observations and date span per survey per object is highlighted in Table \ref{tab:obs}. The full list of observations for all three datasets is listed in Appendix~\ref{ap:A} for 103P and Appendix~\ref{ap:B} for Chiron. In all cases, our most numerous dataset across all respective filters is that of ATLAS ($N_{\rm 103P}=$454, $N_{\rm Chiron}$=1834), followed by ZTF ($N_{\rm 103P}=$242, $N_{\rm Chiron}$=283), then LCO ($N_{\rm 103P}=$149, $N_{\rm Chiron}$=338).

\begin{deluxetable}{c | c c c c c c}
    \tablecaption{Observation date range for our 103P and Chiron datasets, broken down by survey/telescope and filter, including the number of observations $N_{obs}$ obtained in each set, and the number of unique observing nights $N_{nights}$. \label{tab:obs}}
    \tablewidth{\columnwidth}
    \tablehead{
        Object & Survey/Telescope & Filter & UT Start Date   & UT End Date     & $N_{obs}$ & $N_{nights}$
    }
    \startdata
    \multirow{6}{*}{103P/Hartley 2}    & \multirow{2}{*}{ATLAS}    & $c$        & 2023 May 28 & 2024 Mar 31 & 134 & 36 \\
                                       &                           & $o$        & 2023 Jun 6  & 2024 Apr 24 & 320 & 93 \\ \cline{2-7}
                                       & \multirow{2}{*}{ZTF}      & ZTF-$g$    & 2023 Jun 8  & 2024 May 27 & 118 & 67 \\
                                       &                           & ZTF-$r$    & 2023 Jun 6  & 2024 May 03 & 124 & 74 \\ \cline{2-7}
                                       & \multirow{2}{*}{LCO/LOOK} & SDSS $g'$  & 2023 Sep 24 & 2024 Jan 23 & 72  & 65 \\
                                       &                           & SDSS $r'$  & 2023 Sep 24 & 2024 Jan 24 & 77  & 71 \\
    \hline
    \multirow{6}{*}{(2060) Chiron}     & \multirow{2}{*}{ATLAS}    & $c$        & 2020 Jul 20 & 2026 Jan 13 & 421  & 121 \\
                                       &                           & $o$        & 2020 Jun 24 & 2026 Jan 10 & 1413 & 402 \\ \cline{2-7}
                                       & \multirow{2}{*}{ZTF}      & ZTF-$g$    & 2020 Aug 20 & 2026 Jan 20 & 116  & 109  \\
                                       &                           & ZTF-$r$    & 2020 Jul 4  & 2026 Jan 20 & 167  & 166 \\ \cline{2-7}
                                       & \multirow{2}{*}{LCO/LOOK} & SDSS $g'$  & 2022 Sep 5  & 2025 Dec 10 & 166  & 76  \\
                                       &                           & SDSS $r'$  & 2021 Sep 6  & 2025 Dec 10 & 172  & 79  \\ 
    \enddata
\end{deluxetable}

\begin{figure}
    \centering
    \includegraphics[width=\linewidth]{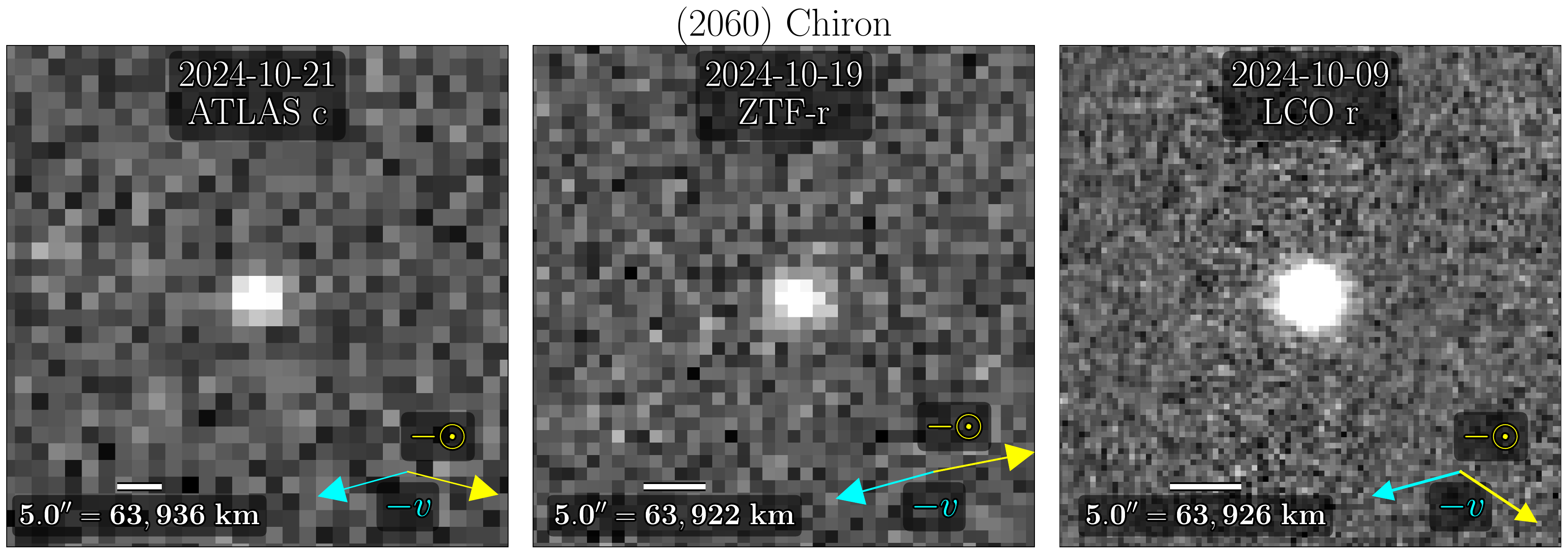}
    \caption{Example observations of (2060) Chiron for survey/telescopes (\textit{left}) ATLAS, (\textit{middle}) ZTF, and (\textit{right}) LCO/LOOK. On all images, the scale of the aperture size used is demarcated by a horizontal white line of length 5$\arcsec$. All panels have been rotated so that North is up and East is left. Also marked on each image are the anti-solar vector (yellow), and anti-motion vector (cyan) as obtained from JPL Horizons. On the pixel scale of all images, Chiron does not display any visual degree of extension, even surrounding its maximum brightness period at opposition around UT 2024 October 17.}
    \label{fig:chiron_obs}
\end{figure}

\begin{figure}
    \centering
    \includegraphics[width=\linewidth]{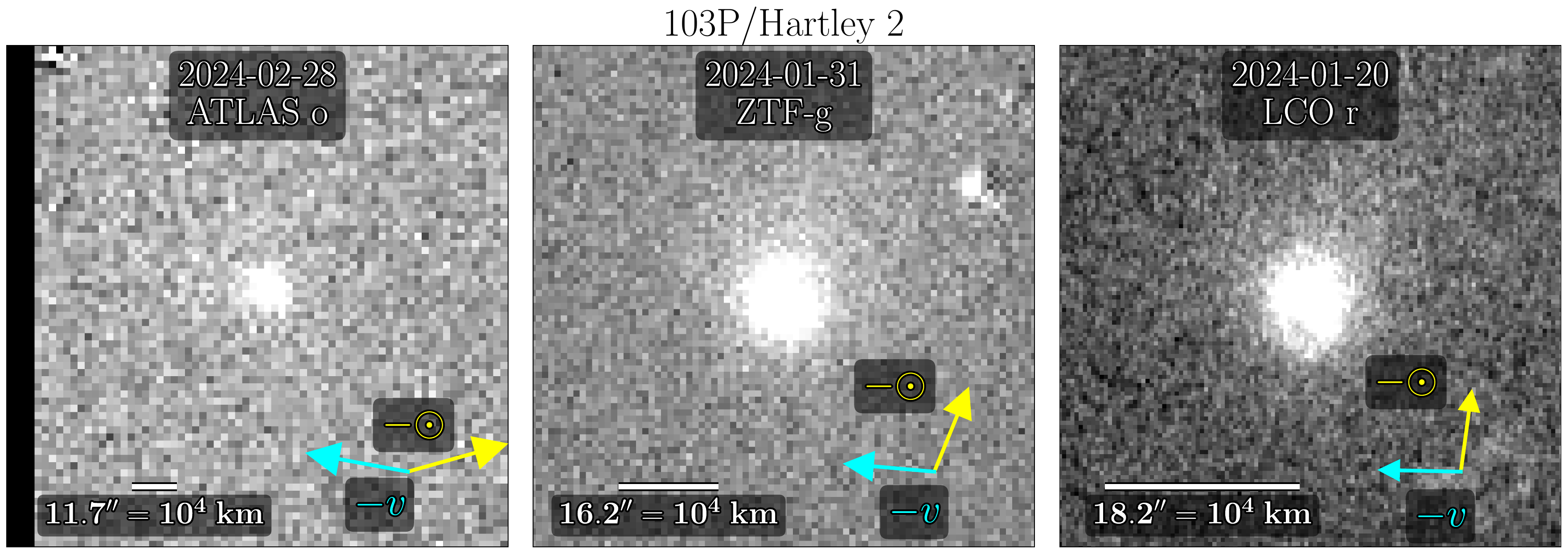}
    \caption{Example observations of 103P/Hartley 2 for survey/telescopes (\textit{left}) ATLAS, (\textit{middle}) ZTF, and (\textit{right}) LCO/LOOK. On all images, the scale of the aperture size used is demarcated by a horizontal white line of length 10,000 km. All panels have been rotated so that North is up and East is left. Also marked on each image are the anti-solar vector (yellow), and anti-motion vector (cyan) as obtained from JPL Horizons. On all images there is a varying degree of extension as 103P moves away from its perihelion passage date of 2023 October 12.}
    \label{fig:103p_obs}
\end{figure}

\subsection{ATLAS} \label{sec:2.1}

ATLAS is made up of four 0.5 m Wright-Schmidt telescopes, with two in Hawai'i (ATLAS-HKO and ATLAS-MLO, commenced in 2015 and 2017 respectively), one in Chile (ATLAS-CHL, commenced in 2022), and one in South Africa (ATLAS-SAAO, commenced in 2022) — one additional site composed of 16$\times$0.3~m telescopes has recently been installed in Tenerife (ATLAS-TDO, commenced 2025). Each are capable of covering 28.9 deg$^2$ field of view in a given exposure \citep{tonry18}. The first four ATLAS units observe in two wideband filters, cyan (or $c$, spanning 420-650 nm), and orange (or $o$, spanning 560-820 nm), covering the majority of the Sloan $gri$ wavelength range to optimize for faint object flux and rapid sky coverage to a limiting depth of $\sim19.5$ mag. Each ATLAS unit revisits the same field four times per night, tiling the entire night sky each night \citep{tonry18}. The ATLAS data reduction pipeline is described in \cite{tonry18, tonry18b}, and \cite{smith20}; observations are astrometrically and photometrically calibrated using the ATLAS RefCat2 catalog \citep[with astrometry derived from Gaia,][]{gaia18, tonry18}, whereafter difference imaging is performed, removing static stars and galaxies. 

Our photometry for Chiron over the 2020-2026 observing period is obtained from serendipitous sidereally tracked observations via the ATLAS Forced Photometry Server \citep{shingles21}, whereby a PSF is fit at predicted object coordinates from orbital ephemerides supplied by the MPC. AB magnitudes are calculated from the flux at this position — we opt to utilize difference images for this calculation rather than reduced images in order to reduce flux contamination from nearby sources. This approach is well-suited for Chiron where, even during peak activity in 2021 no resolvable coma was observable, instead showing a radial profile consistent with background field stars \citep{dobson24}. This trend continues into opposition during its 2024 season, as seen by the lack of visual extension in Figure \ref{fig:chiron_obs}. As a result, Chiron appears effectively point-like at the $\sim1.86\arcsec$/pix resolution of ATLAS \citep{tonry18}. Observations are then filtered to ensure good quality detections of Chiron by checking that the ATLAS-calculated magnitude is (i) brighter than the 5$\sigma$ limiting magnitude of the image and (ii) brighter than the $3\sigma$ upper-magnitude limit from the uncertainty on the measured flux. 

Conversely, visual inspection of images shows that 103P displays a varying degree of extension over its 2023/24 apparition, as showcased in Figure \ref{fig:103p_obs}. Therefore, we measure its flux using aperture photometry on the difference images. We adopt a circular aperture with radius corresponding to a projected 10,000 km, a scale commonly utilized in literature cometary studies for inter-comparison \citep[e.g.][]{furusho07, opitom15, snodgrass16, moreno16, ivanova19, moulane23, gillan24, gillan25}. This aperture is centered on the centroided position of the comet's predicted coordinates from JPL Horizons\footnote{\href{https://ssd.jpl.nasa.gov}{https://ssd.jpl.nasa.gov}}. Local background within the aperture was subtracted using concentric sky annuli placed immediately outside of the aperture, calculating a total background from the product of the area of the annulus and the 3$\sigma$-clipped median background value within it. This approach ensures that both nuclear and inner-coma contributions were consistently captured across all observations with minimal field star contamination. 

\subsection{ZTF} \label{sec:2.2}

ZTF utilizes the 1.2 m Samuel Oschin Schmidt Telescope at the Palomar Observatory, with a 47 deg$^2$ field of view camera composed of sixteen 6k$\times$6k CCDs, yielding a pixel scale of $1.01\arcsec$/pix \citep{bellm19, graham19, dekany20}. ZTF observes in three broad-band filters, ZTF-$g$ (spanning $\sim367-561$ nm) ZTF-$r$ (spanning $\sim550-739$ nm), and ZTF-$i$ (spanning $\sim687-986$ nm) — these filters differ from the SDSS-$g/r/i$ bandpasses in that they display relatively steep transmission edges with a box-like profile, providing high throughput and well-defined band limits designed to avoid some bright sky lines at Palomar \citep{bellm19}. Since 2018, ZTF has covered $\gtrsim25,000$ deg$^2$ of the visible night sky north of a declination of $\delta\approx-31^\circ$ every two to four nights, to a 5$\sigma$ limiting depth in ZTF-$r$ of $\sim20.6$ mag \citep{masci19, dekany20}. The ZTF processing procedure is described in \cite{masci19}; images are astrometrically calibrated with the Gaia DR2 catalog \citep{gaia18}, and photometrically via the Pan-STARRS1 DR1 catalog \citep{chambers16}. 

Our Chiron photometry has been obtained from serendipitous sidereally tracked observations via the community broker for ZTF solar system object alerts, FINK \citep{moller21}. Photometry is obtained from PSF fitting of sources that have been tagged as solar system objects by the ZTF pipeline. As before however, 103P photometry is measured using aperture photometry with a 10,000~km radius aperture on the predicted JPL Horizons ephemerides — all employed via the \texttt{ZChecker} tool \citep{kelley19}. Data are filtered to ensure quality, removing observations with measured magnitude uncertainties $\sigma_m \gtrsim 0.36$ mag, and manually inspecting for outliers (e.g.\ image processing artifacts, poor centroiding, contaminating stars).

\subsection{LCO} \label{sec:2.3}

The LCO operates a globally distributed network of 25 robotic telescopes, ranging in aperture size from $0.4-2.0$ m \citep{brown13}. Facilities are located in sites in both hemispheres in order to enable near-spatially-continuous sky coverage. For the LOOK Project, we programatically scheduled observations in the SDSS $g'$ (spanning $\sim386-569$ nm) and SDSS $r'$ (spanning $\sim541-718$ nm) filters with the 1.0 m telescopes via the NEOExchange web portal \citep{lister21}. Each 1.0 m is equipped with a Sinistro imager, made up of a 4096$\times$4096 pixel CCD, yielding a $26.6^\prime \times 26.6^\prime$ field of view with a 0.387$\arcsec$/pix scale \citep{brown13}, whilst the 2.0 m telescopes utilize the Multicolor Simultaneous Camera for studying Atmospheres of Transiting exoplanets (MuSCAT) instrument, with its four $2048\times2048$ pixel CCDs, each having a field of view of $9.1^\prime \times 9.1^\prime$, with a pixel scale of 0.27$\arcsec$/pix \citep{narita20}. Observations from all telescopes are reduced using the Beautiful Algorithm to Normalize Zillions of Astronomical Images \citep[BANZAI,][]{mccully18} pipeline, then calibrated to the Pan-STARRS1 system using the ATLAS RefCat2 photometric catalog \citep{tonry18b}, the \texttt{CALVIACAT} photometric calibration software \citep{kelley21}, and BANZAI measured field stars \citep{lister22}. 

Observations of Chiron are obtained using non-sidereal tracking at half the object's apparent motion, meaning that the telescope tracks at half the ephemeris rate so that both the target and background stars are trailed by the same amount. In contrast, observations of 103P are taken using full non-sidereal tracking at the object's ephemeris rate, leaving the comet untrailed, whilst background stars are elongated. Chiron is measured utilizing a 5$\arcsec$ radius to better sample the PSF than a cometary 10,000 km aperture, which at Chiron's average geocentric distance of $\sim18.1$ au and the 1.0 m 0.387$\arcsec$/pix scale corresponds to $\sim0.76\arcsec$, which is much less than the typical $\sim2\farcs3$ seeing of the observations. For 103P we use a standard 10,000 km radius (corresponding to $\sim6-36\arcsec$ over our time range) to further allow inter-comparison. All observations are visually inspected for outliers and filtered to those with seeing $\lesssim5.0\arcsec$ to ensure accurate measurement of object flux.

\section{Data Processing \& Calibration} \label{sec:3}

For all of our observations from all surveys/telescopes, we first correct for light travel time, and then correct our apparent magnitudes $m_\lambda$ in each filter $\lambda$ to reduced magnitudes (the apparent magnitude an object would have at simultaneously a heliocentric and geocentric distance of 1 au) by the following equation:

\begin{equation}
    M_\lambda(1, 1, \alpha) = m_\lambda - 5 \log_{10}(r_\mathrm{H} \Delta)
\end{equation}
where the heliocentric distances $r_\mathrm{H}$ and geocentric distances $\Delta$ are obtained instantaneously from JPL Horizons\footnote{\href{https://ssd.jpl.nasa.gov}{https://ssd.jpl.nasa.gov}}. When required, we additionally apply a phase correction to report phase-normalized reduced magnitudes:

\begin{equation}
    M_\lambda(1, 1, 0) = M_\lambda(1, 1, \alpha) + 2.5\log_{10}(\Phi(\alpha))
\end{equation}
where $\Phi(\alpha)$ is a phase function governed by the Sun-target-observer (phase) angle $\alpha$ of the observation. For 103P, we apply the empirical, standardized Schleicher-Marcus dust phase function \citep{schleicher98, marcus07}\footnote{Accessed at \href{https://asteroid.lowell.edu/comet/dustphase/}{https://asteroid.lowell.edu/comet/dustphase/}} to correct the $r$-band photometry to $0^\circ$ phase angle, thereby accounting for dust scattering in the coma. In contrast, the $g$-band signal contains significant gaseous emission (primarily C$_2$). Unlike dust, the gas emission has minimal variation with phase angle \citep{ahearn95, schleicher04} — thus we elect not to apply any phase correction to $g$, since the relative contributions of gas and dust to the brightness are not known \textit{a priori}. The $g$ signal is therefore used as a crude tracer of volatile activity. As Chiron has exhibited an evolving phase curve slope since its 2021 outburst \citep{dobson24}, we do not correct for phase angle here, instead measuring each epoch individually to probe changes (Section \ref{sec:4.2.3}).

Each filter of each dataset for each object is treated individually. We visually inspect for obvious outliers in our dataset due to e.g.\ contaminating flux from background stars or non-detection in the photometry pipelines described in Section \ref{sec:2}. We then apply a rolling-window sigma-clipping procedure, where each data point is compared against the local distribution of its neighbors within an 11 day window (i.e.\ 5 days either side of a given data point), and outliers that deviate beyond a 3$\sigma$ threshold are flagged and removed — in practice, this removes spurious measurements from the dataset, ensuring that the sample more accurately represents the underlying trend in brightness evolution. Finally, we remove data points whose magnitude uncertainty measurements lay outside of the 85$^{\text{th}}$ percentile of the magnitude uncertainty distribution\update{, corresponding to $\sigma{_{m,ZTF}} \gtrsim$ 0.08~mag and $\sigma{_{m,ATLAS}} \gtrsim$ 0.15~mag}.

In order to obtain a more uniform dataset, we color-shift our photometry onto a common reference filter. For Chiron, we shift the ZTF-$g$ and LCO/LOOK $g$ photometry to the more densely sampled ATLAS $c$ filter, and similarly the ZTF-$r$ and LCO/LOOK $r$ to ATLAS $o$ filter. For 103P, we instead shift the ATLAS photometry to the ZTF filter system, as ZTF provides denser sampling during the peak activity phase. To do this, we follow a similar method to \cite{dobson24} and fit third-order polynomial splines to all datasets via the \texttt{UnivariateSpline} routine from \texttt{scipy.interpolate} \citep{virtanen20}; for 103P we split the observing epoch into pre-, during-, and post-peak activity splines to more accurately capture the shape of the lightcurve and potentially identify any sudden changes in brightness due to outbursts, whilst for Chiron we fit splines per observing season. For a given point on an object's light curve, we can then calculate a corresponding magnitude at the same time on the other splined light curve, allowing us to calculate filter shifts (e.g.\ $g-c$ and $r-o$) for every data point. We only perform this method for regions where $g/c$ and $r/o$ overlap temporally, as extrapolating beyond this in a given filter can produce poor spline fits to the dataset and lead to inaccurate spline magnitude estimates. The spline fits to each dataset for each object are shown in Sections \ref{sec:4.1.1} and \ref{sec:4.2.1}.

\section{Results} \label{sec:4}
With our observations from ATLAS, ZTF, and LCO we are able to analyze the light curve and phase curve evolution, activity searches, mass-loss rates, color evolution, and rotational period searches for both 103P and Chiron. In this section we analyze each separately, discussing the implications that the observed trends in both photometric datasets have for the long-term behavior of both objects.

\subsection{103P/Hartley 2} \label{sec:4.1}

\subsubsection{Light Curve Evolution} \label{sec:4.1.1}

\begin{figure}
    \centering
    \includegraphics[width=0.9\linewidth]{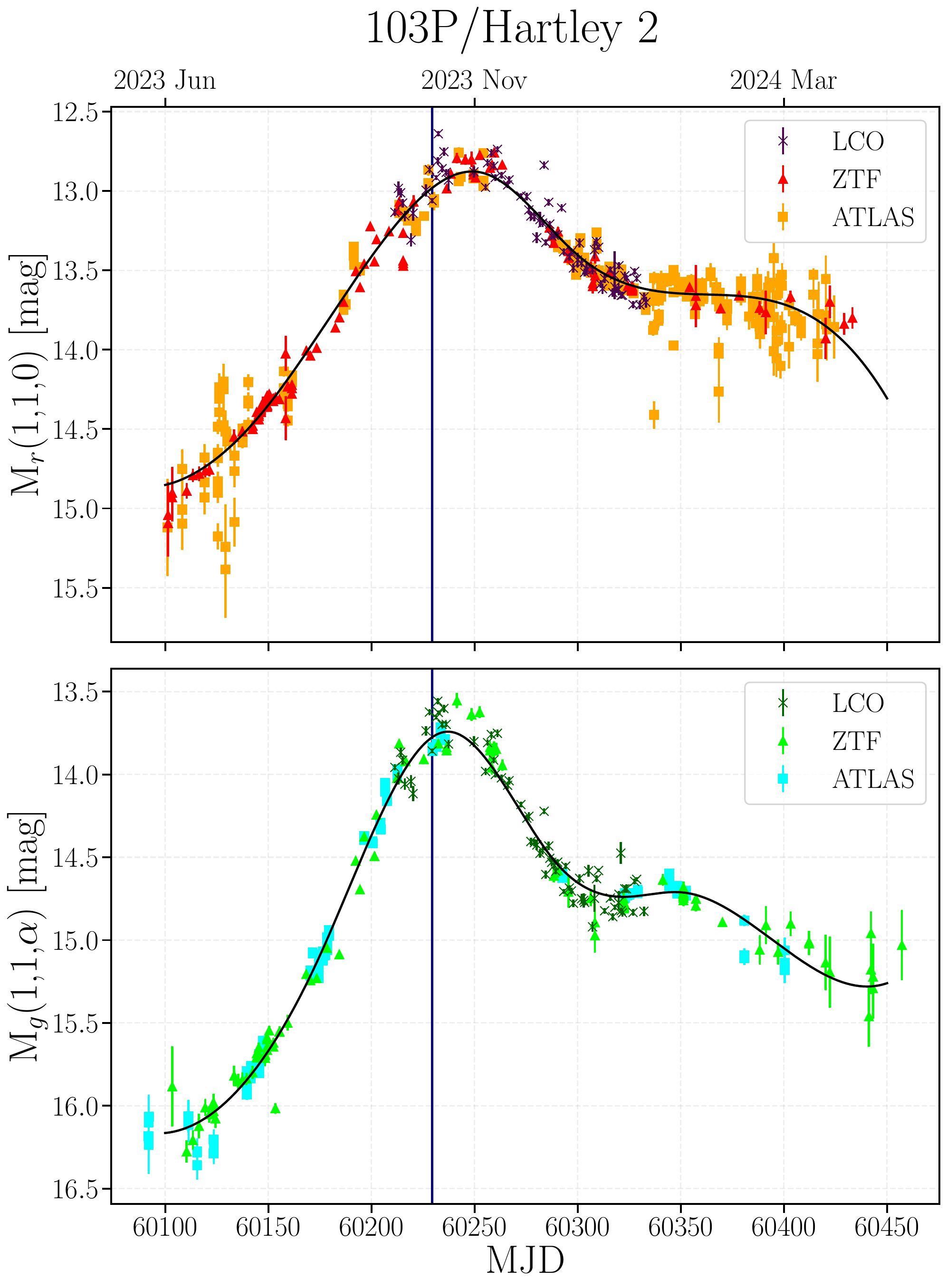}
    \caption{\update{Distance- and phase-corrected light curves for 103P/Hartley 2 color-shifted to ($top$) ZTF-$r$ band and ($bottom$) to ZTF-$g$ band. Note that no phase correction is applied to the $g$-band (see Section~\ref{sec:3}).} The different datasets are denoted via differing colors and markers. The solid black \update{curve} running over each dataset is a third order polynomial spline fit. Overplotted in a solid navy line in both plots is the date of perihelion passage, 2023 October 12, for reference.}
    \label{chap4:fig:103p_lightcurve}
\end{figure}

The distance-corrected (and for $r$-band phase-corrected) light curves of 103P during the 2023/24 apparition are shown in Figure \ref{chap4:fig:103p_lightcurve} after color shifting to ZTF-$g$ and $r$ bands. The brightness displays a smooth rise to a peak brightness of $M_g(1,1,\alpha) = 13.55 \pm 0.04$ mag and $M_r(1,1,\alpha) = 12.64 \pm 0.02$ mag, followed by a slower decline with no outbursts (sudden increase in brightness followed by a quasi-exponential decay) detected on visual inspection. Fitting the heliocentric dependence of the flux as $F \propto r_\mathrm{H}^{n}$ yields a pre-perihelion slope of $n_r = -3.48 \pm 0.08$. This value is consistent with the dust-dominated slopes of \update{$n \approx -2.8$ to $-3.6$} reported by \citet{knight13} for the 2010/11 apparition \update{at comparable heliocentric range of $r_\mathrm{H} \sim$ 1.68 - 1.06~au}. After perihelion, the trend reverses: $n_r = -1.16 \pm 0.04$, somewhat shallower than the \update{$n \approx -1.7$ to $-1.9$} measured by \citet{knight13}. As in 2010/11, this flattening can be attributed to the persistence of larger, slower-moving dust grains in the coma\update{: larger grains are less susceptible to radiation pressure and therefore linger within the photometric aperture for longer timescales, continuing to contribute to the measured flux even as volatile-driven activity wanes.} Our results suggest the effect is now more pronounced than in 2010/11, indicating an increasing relative dust contribution to the optical flux post-perihelion. \update{The near-flat region visible in both panels at approximately +100~days post-perihelion ($\sim$2024 Jan) coincides with a minimum in the phase angle at $\sim16^\circ$. For the uncorrected $g$-band data, this geometric effect will partially counteract the declining coma brightness, as reduced phase angles suppress the forward scattering losses. However, the leveling-off is also present in the phase-corrected $r$-band, which implies that the phase angle geometry alone is insufficient to explain the feature, and that an intrinsic slowing of the post-perihelion activity decline at this epoch may be contributing.} Combined with the secular fading reported by \citet{graykowski25}, the light-curve behavior seen here is consistent with weakening overall activity, although the volatile component cannot be exclusively constrained without dedicated narrowband/spectroscopic gas measurements.

\subsubsection{Mass-loss Estimates}
We estimate the dust mass-loss rate by first subtracting a nucleus-only baseline from the nuclear absolute magnitude. We adopt $H_{r,\mathrm{nuc}}=18.2$, converted from the EPOXI-measured $H_{V,nuc}=18.4$ mag \citep{li13}, using the Johnson-SDSS transformation \citep{jordi06}. This isolates an approximation of a ``coma-only'' light curve. We assume a steady-state dust coma and convert the total scattering cross-section $C$ into a dust production rate $\sfrac{dM}{dt}$ by the following equation:

\begin{equation}
    \frac{dM}{dt} = \frac{8}{3} \rho a \frac{C}{\tau}
\end{equation}
where $\rho$ is the grain density and $a$ is the particle radius. The residency time $\tau$ within the aperture of dust grains with a velocity $v$ is given as $\tau = \sfrac{10,000~\text{km}}{v}$, and the scattering cross-section $C$ in m$^2$ can be found by:

\begin{equation}
    pC = (2.25\times10^{22})\pi 10^{-0.4(M(1,1,0)-M_\odot)}
\end{equation}
where $p$ is the geometric albedo of the grains, and $M_{\odot,r} = -26.93$ is the solar magnitude \citep{willmer18}. We first evaluate a small-particle case, motivated by thermal-infrared modeling which indicates a coma size distribution strongly peaking at $\sim0.5-0.9$~$\mu$m, consistent with dark \citep[$p=0.04$,][]{li13, harker18} sub-micron grains with $\rho=1$~g~cm$^{-3}$ outflowing with $v=100$~m~s$^{-1}$ dominating the cross-section. This yields a peak dust mass-loss rate of $\lesssim 4$~kg~s$^{-1}$, which is negligible compared to EPOXI measured production rates of H$_2$O and CO$_2$ \citep[$\sim300$ and $\sim160$~kg~s$^{-1}$ respectively,][]{ahearn11}. 

However, EPOXI measurements also reveal a population of cm-scale icy grains in 103P's coma, comprising $\sim16\%$ of the scattering cross-section within 21~km of the nucleus \citep{kelley13, kelley15}. Adopting $a=1$~cm, $p=0.31$ \citep{russell16, kelley15}, and $v=1$~m~s$^{-1}$ \citep{hermalyn13} for this sub-population, their contribution to the peak dust mass-loss rate is $\lesssim12$~kg~s$^{-1}$. Although large grains represent a minority of the scattering cross-section, their much larger grain radius means they dominate the mass budget. Their relative contribution is also expected to evolve with distance from the nucleus, as cm-scale icy grains sublimate on comparatively shorter timescales \citep{markkanen20} and exhibit stronger tailward-biased dynamics, unlike the more spatially extended fine-grained dust \citep{kelley13, mueller13}. Our combined estimate of $\lesssim16$~kg~s$^{-1}$ is consistent with the ATLAS-$o$-band upper limit of \cite{gillan25}\update{, who obtain $\lesssim16$~kg~s$^{-1}$} \update{by performing aperture photometry on ATLAS-$o$ data with a 10$^4$~km aperture,} and assuming \update{dark,} micron-sized grains \update{with high outflow velocities} dominating the cross-section\footnote{\cite{gillan25} report $\lesssim63$~kg~s$^{-1}$ for 103P in their Table~2, however this was calculated using the ATLAS-$c$ band photometry rather than ATLAS-$o$ band (private communication). \update{The updated value of $\lesssim16$~kg~s$^{-1}$ was obtained by applying their photometry method to their dataset as intended for this analysis.}}. The agreement is physically reasonable since small grains dominate the cross-section regardless, with the large-grain mass contribution effectively absorbed into their faster assumed outflow velocity. \update{All such estimates, including those of \cite{gillan25}, are highly model-dependent upper limits and should not be interpreted as precise measurements;} these are best treated as order-of-magnitude, bracketing estimates of activity strength.

\subsubsection{Color Evolution} \label{sec:4.1.2}
The dense, dual-filter coverage between ATLAS, ZTF, and LCO allows us to measure changes in 103P's ($g-r$) colors over its apparition. We employ the two complementary approaches \update{of} \cite{dobson23, dobson24}\update{: a nearest-neighbor method pairing each observation with the temporally closest observation in the second filter within $\pm$2~days, and a spline-based Monte Carlo approach fitting $10^5$ synthetic light curves per filter to interpolate colors at shared epochs.} Both methods \update{use} magnitudes corrected for heliocentric and geocentric distance but not phase angle ($M(1,1,\alpha)$)\update{ — as dust-scattered light exhibits a phase dependence whilst molecular emission is largely isotropic, applying a dust phase correction to the likely gas-contaminated $g$-band may introduce a phase-dependent systematic into ($g-r$) color estimates. A rigorous decomposition of dust continuum and gas emission would be strongly model-dependent given the mixed bandpasses and cadences of this dataset, and is beyond the scope of this work.}

\begin{figure}
    \centering
    \includegraphics[width=\linewidth]{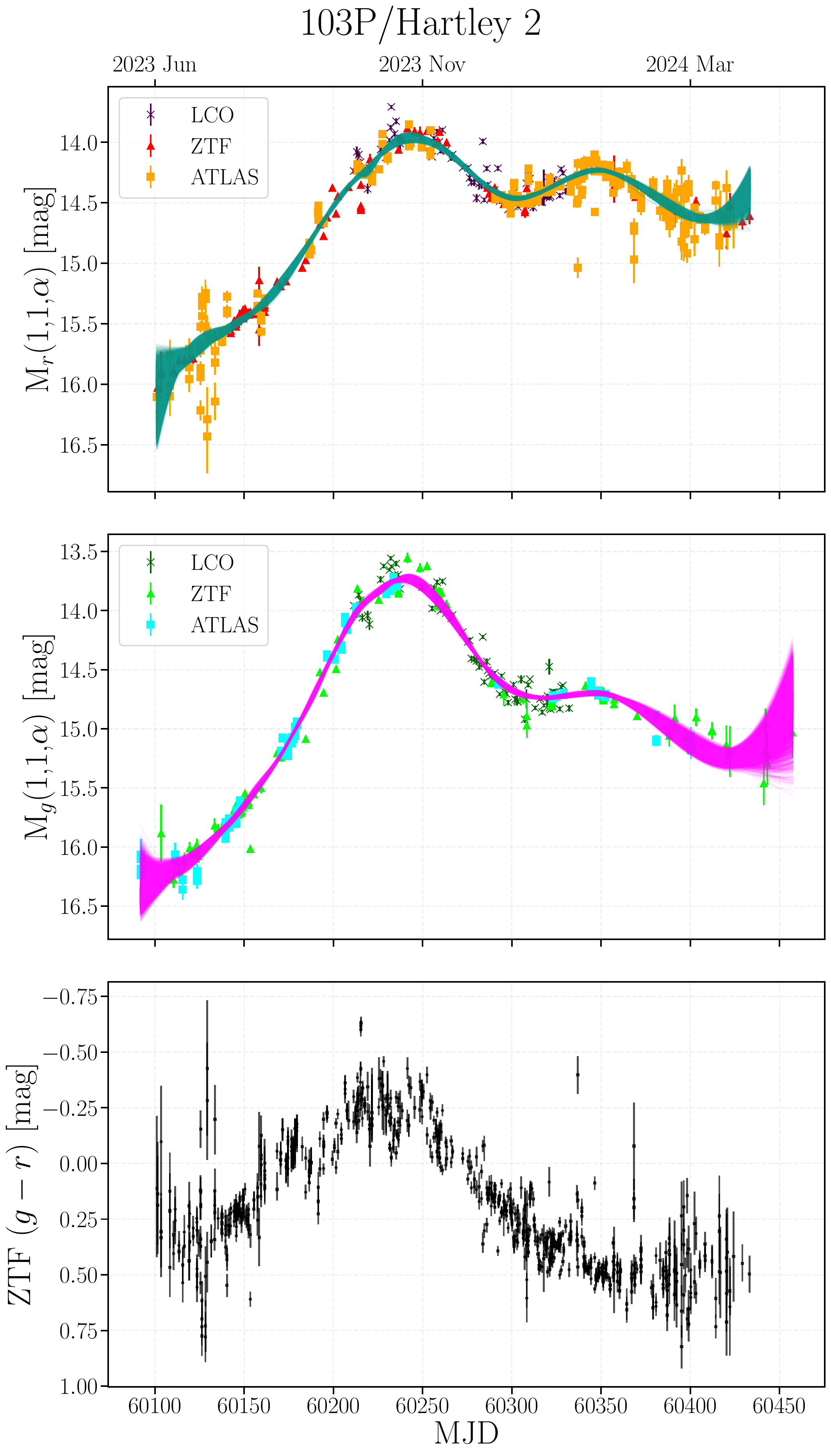}
    \caption{Light curves of the color-shifted magnitudes of 103P to ($top$) the ZTF-$r$ band and ($middle$) the ZTF-$g$ band. Overplotted on each are $10^5$ synthetic Monte Carlo sampled splines within measurement uncertainties. ($bottom$) The color evolution of 103P in ZTF ($g-r$) over the shared time range, with uncertainty on each data point being estimated by the 1$\sigma$ uncertainties from the splines.}
    \label{fig:103p_colevo_spl}
\end{figure}

\begin{figure}
    \centering
    \includegraphics[width=\linewidth]{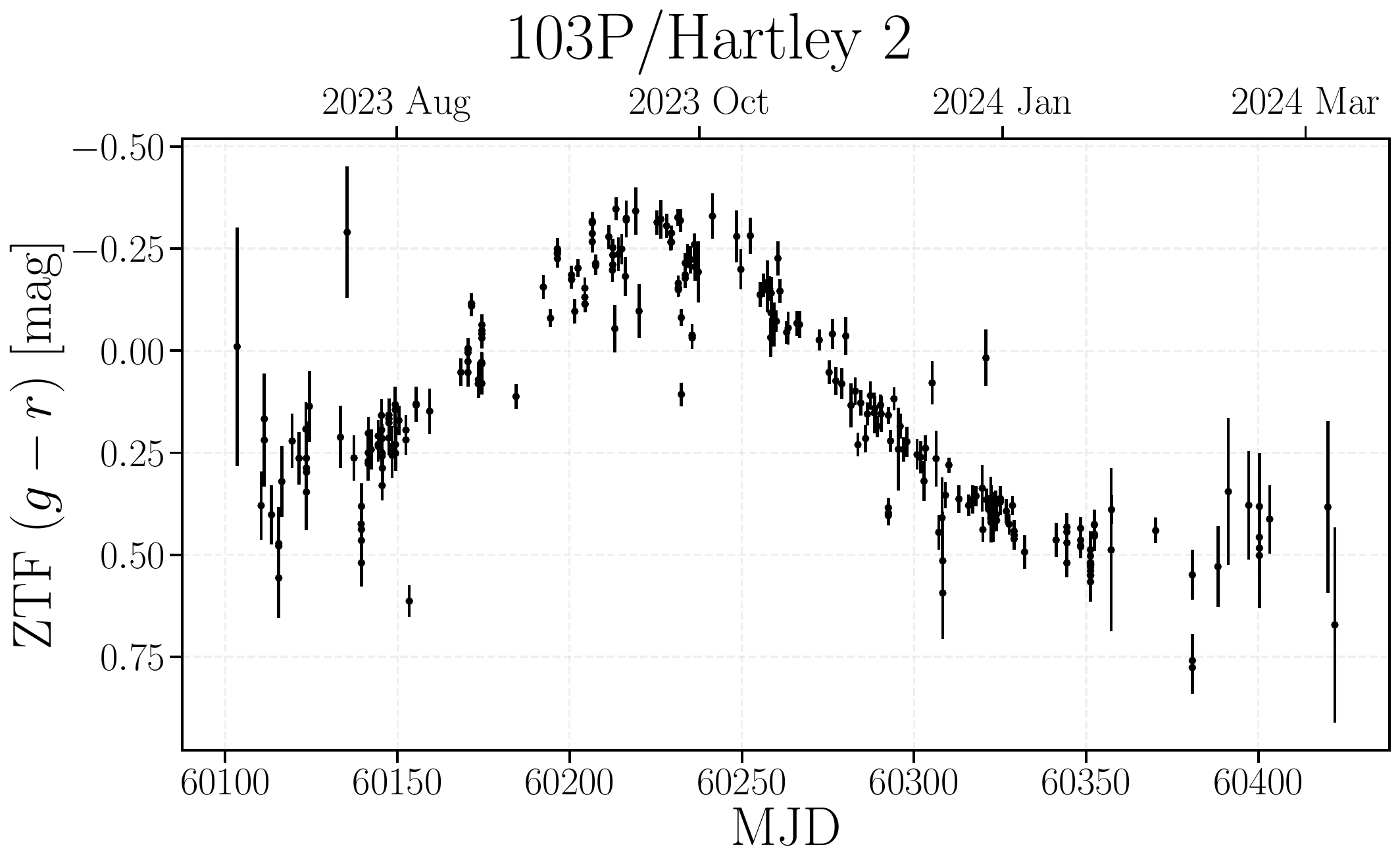}
    \caption{Color evolution of 103P in ZTF ($g-r$) as in Figure~\ref{fig:103p_colevo_spl}, this time using a nearest-neighbor approach with a time separation of $\pm$2 days. Uncertainties are obtained from propagating the individual photometric uncertainties in quadrature.}
    \label{fig:103p_colevo_nn}
\end{figure}

The ($g-r$) color of 103P shows \update{clear but moderate} evolution across the 2023/24 apparition, \update{as shown in Figures~\ref{fig:103p_colevo_spl} and \ref{fig:103p_colevo_nn}.} Prior to perihelion, the coma was slightly redder-than-solar \citep[($g-r$)$_{\odot}=0.39$,][]{willmer18}, consistent with dust-dominated continuum scattering. \update{Near perihelion,} both methods reveal a pronounced blueward excursion\update{, reaching $(g-r)\sim-0.2$ mag, an amplitude difficult to explain from dust-color changes alone. This is instead more} consistent with an increased fractional contribution from molecular emission within the $g$ band (most plausibly C$_2$; \citealt{ahearn11, knight13, protopapa14})\update{, supported by the broad and nearly isotropic coma component visible in $g$-band morphology.} Following perihelion, \update{the color returned to $\sim0.5$ mag as activity waned, consistent with declining gas emission and growing dust dominance --- behavior that is also seen in long-period comets \citep{holt24}, suggesting bluer $g-r$ at smaller heliocentric distances is a common signature of increasing gas emission.}

\update{We emphasize that these color variations} do not uniquely constrain changes in the dust-to-gas production ratio\update{: even at constant production ratios, daughter species such as C$_2$ may exhibit a steeper heliocentric distance dependence than dust due to fluorescence efficiency and photochemical production rates being governed by solar irradiation \citep{cochran85}.} The coincidence of the bluest colors with the period of peak photometric activity \update{(Section~\ref{sec:4.1.1}) suggests} that molecular emission most strongly influenced the observed \update{colors} near perihelion. Overall, the apparition \update{shows dust-dominated colors at large} heliocentric distance, a transition to strongly gas-contaminated $g$-band photometry near perihelion, and a gradual return \update{to} dust-dominated conditions — consistent with prior \update{studies} of 103P \citep{meech11, knight13, reach13, kelley15}.

\subsubsection{Rotational Period Search} \label{sec:4.1.3}

\citet{lehmann25} recently reported that the rotational period of 103P is lengthening at a rate of $\sim20$~s~day$^{-1}$, slower than the $\sim60$~s~day$^{-1}$ measured by \citet{knight15}, suggesting a reduced activity-driven torque and potentially consistent with the long-term activity decline noted by \citet{graykowski25}. \update{Here we assess the extent to which our sparse, broadband survey photometry can independently recover and constrain such rotational variability, both as an individual check and as a test of what survey datasets can realistically contribute to cometary rotation studies.}

\update{We subtracted spline fits to the color-shifted ZTF-$g$ and ZTF-$r$ light curves to isolate rotational residuals, then applied the Generalized Lomb-Scargle (GLS) periodogram \citep{lomb76, scargle82, zechmeister09}, scanning 10$^3$ frequencies over 13-23~hr, covering the plausible rotation period range. We analyzed the $g$ and $r$ bands separately, as rotational variability in 103P is known to be more readily detectable in gas-sensitive bands where emitting gas moves rapidly through the photometric aperture, producing higher-contrast variability than the more slowly evolving dust continuum \citep{knight11, knight13, knight15}. To quantify the limits of this approach, injection-recovery tests (sinusoidal signals, amplitudes 0.10-0.30~mags, periods 8-22~hr) confirm that amplitudes $\gtrsim0.15$~mag are recoverable across the full period range, with very short periods ($\lesssim$10~hr) less accessible due to diurnal gaps; white-noise control periodograms show no systematic peaks, confirming that any detected signals are not cadence artifacts.}

\begin{figure*}
    \centering
    \includegraphics[width=\textwidth]{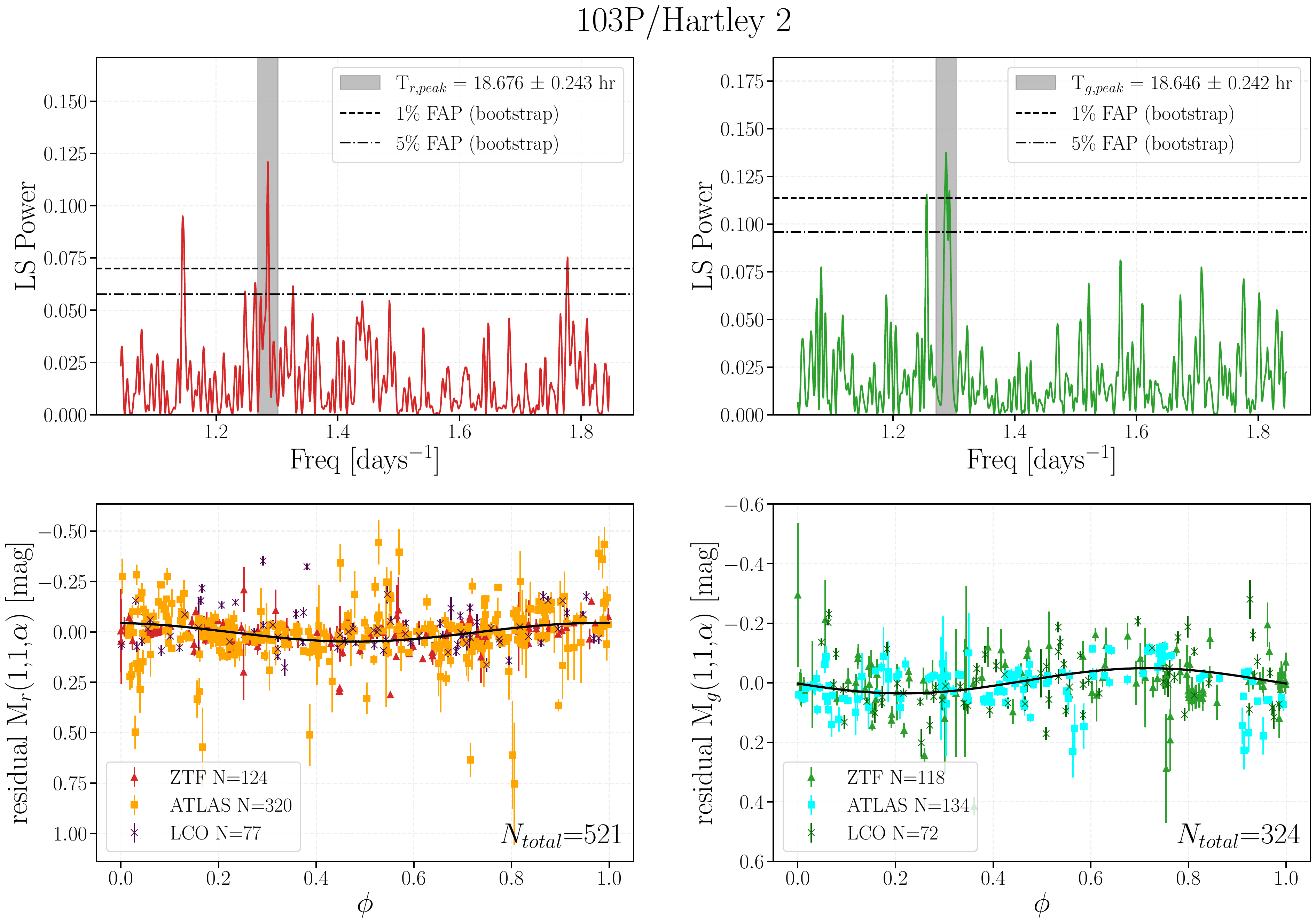}
    \caption{($top$) Generalized Lomb-Scargle periodograms for the measurements color-shifted to ($left$) ZTF $r$-band and ($right$) ZTF $g$-band measurements of 103P. Highlighted in gray is the strongest power frequency peak, with the False Alarm Probability (FAP, the probability that such a peak would arise from noise alone) 1\% and 5\% levels displayed as dashed and dash-dotted lines respectively. ($bottom$) shows the dataset phase-folded $\phi$ on the corresponding best period, with a simple single-peaked sinusoid model overlaid. The legend displays the number of data points used in each survey/telescope to construct the periodogram.}
    \label{fig:103p_period_all}
\end{figure*}

\update{The GLS periodograms in Figure~\ref{fig:103p_period_all} yield dominant peaks at $T_g = 18.65$~hr and $T_r = 18.68$~hr, both exceeding the bootstrapped 1\% false alarm probability (FAP) threshold and consistent with the 18.6-19.1~hr range from \citet{lehmann25}. These values fall at the slower end of the rotational evolution documented during the 2010 apparition, in which \citet{knight15} reported an apparent lengthening from $\sim16$~hr to $\sim19$~hr over a single return. If comparable spin-down had accumulated across subsequent apparitions, periods of $\sim20-22$~hr might be expected by 2023 --- their absence here suggests rotational evolution in 103P does not proceed monotonically between returns.} One possibility is that the activity-driven torque varies between returns due to seasonal illumination effects or evolving active region geometry, as inferred from EPOXI observations \citep{ahearn11, knight11, knight15} and documented in 67P/Churyumov-Gerasimenko \citep{keller15, lai16}. Alternatively, the complex NPA rotation state may produce apparent period variations from the interplay of its component periods, mimicking secular spin evolution within a single apparition whilst yielding similar apparent periods at comparable orbital phase across different returns. Non-linear and cyclic behavior of this nature has also been observed in 67P \cite{keller15}. \update{The NPA state likely contributes directly to the modest GLS powers we measure ($\sim0.12-0.15$), as its evolving multi-periodic light curve structure reduces the coherence of any single recovered signal.}

\update{We also subdivided the dataset into pre-perihelion, near-perihelion ($T_P \pm 40$~days), and post-perihelion intervals. Near perihelion, both bands robustly yielded $\sim18.7$~hr ($T_g = 18.69$~hr, $T_r = 18.73$~hr), consistent with \cite{knight15} and \cite{lehmann25}. Outside of this window, period recovery degrades however: pre-perihelion ZTF-$g$ produced $\sim19.06$~hr whilst ZTF-$r$ gave $14.92$~hr (the latter is inconsistent with any prior measurement and is likely a nightly-cadence alias), whilst post-perihelion sees these disagreements reverse. We therefore regard only the near-perihelion result as a significant, activity-linked periodicity, and do not attempt to derive a spin-down rate from this dataset alone.}

\begin{figure}
    \centering
    \includegraphics[width=\columnwidth]{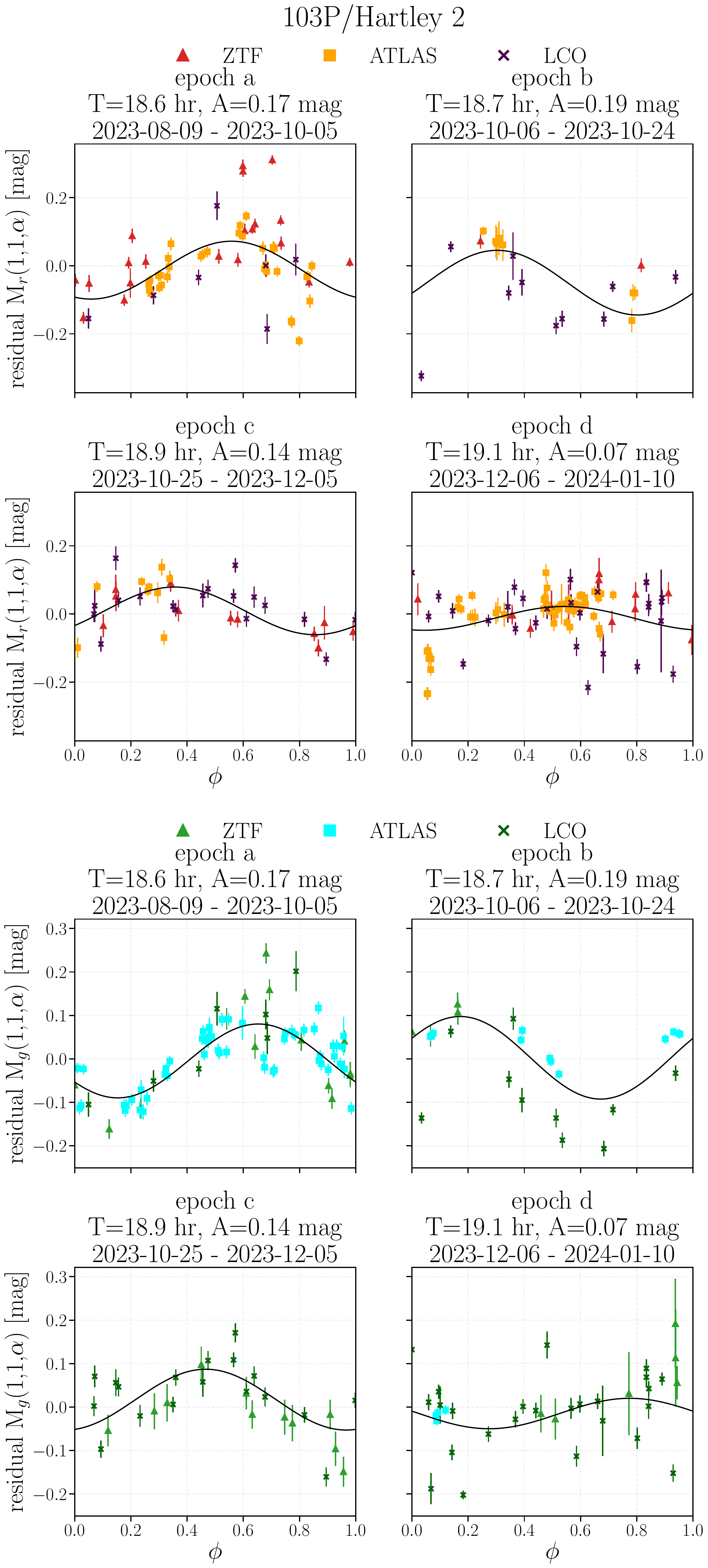}
    \caption{Rotational phase-folded residual light curves for 103P using the same epochs, measured rotational period $T$, and light curve amplitudes $A$ (denoted on top of each subplot) from \cite{lehmann25} with our survey/telescope dataset. ($top$) is the color-shifted to ZTF-$r$ band data, and ($bottom$) is the color-shifted to ZTF-$g$ band data. Overplotted in solid blue is the best fit single-peaked sinusoid measured.}
    \label{chap4:fig:103p_periodogram_thomas}
\end{figure}

As a direct comparison with \citet{lehmann25}, we phase-folded our $g$- and $r$-band light curves on their reported periods for each time interval, shown in Figure~\ref{chap4:fig:103p_periodogram_thomas}. The folded curves remain largely noise-dominated, though the periodic signal is more discernible in $g$-band than $r$-band, consistent with gas emission providing a higher-contrast rotational tracer than dust. \update{Collectively, these results demonstrate that broadband survey photometry is capable of recovering the broad rotational signal near perihelion, but lacks the gas-band sensitivity and temporal resolution of dedicated narrowband campaigns needed to track period changes on weekly timescales.}

\subsection{(2060) Chiron} \label{sec:4.2}

\subsubsection{Light Curve Evolution} \label{sec:4.2.1}
\begin{figure}
    \centering
    \includegraphics[width=\linewidth]{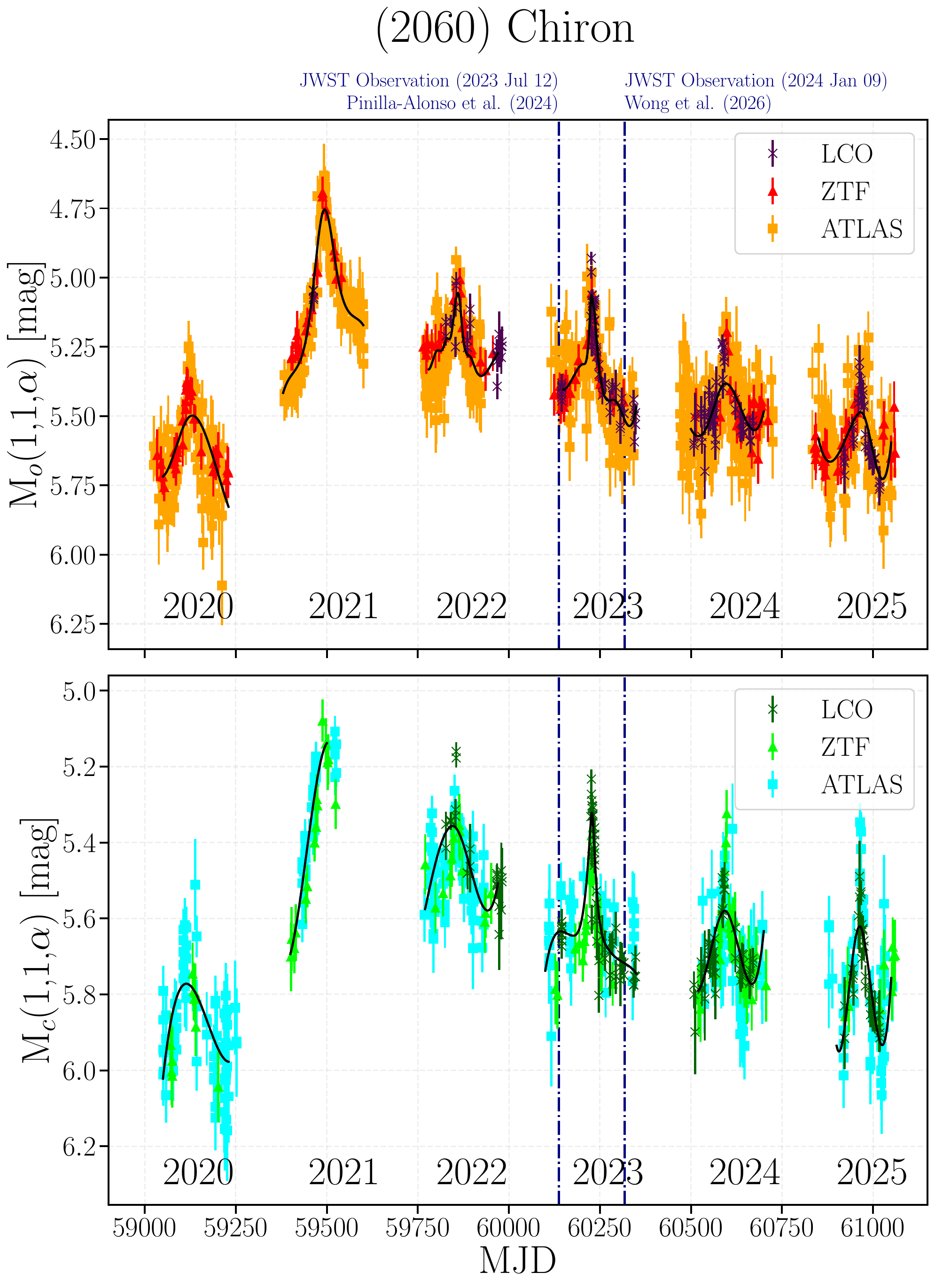}
    \caption{Distance-corrected light curves for (2060) Chiron color-shifted to ($top$) ATLAS $o$ band and ($bottom$) ATLAS $c$ band. The different datasets are denoted via differing colors. The solid black \update{curve} running over each observing season in each dataset is a third order polynomial spline fit. Overplotted in \update{dash dotted navy lines} in both plots is the date of contemporaneous JWST observations on 2023 Jul 12 \citep{pinillaalonso24} \update{and 2024 Jan 09 \citep{wong26}} for reference.}
    \label{chap4:fig:chiron_lc}
\end{figure}

\begin{figure*}
    \centering
    \includegraphics[width=\linewidth]{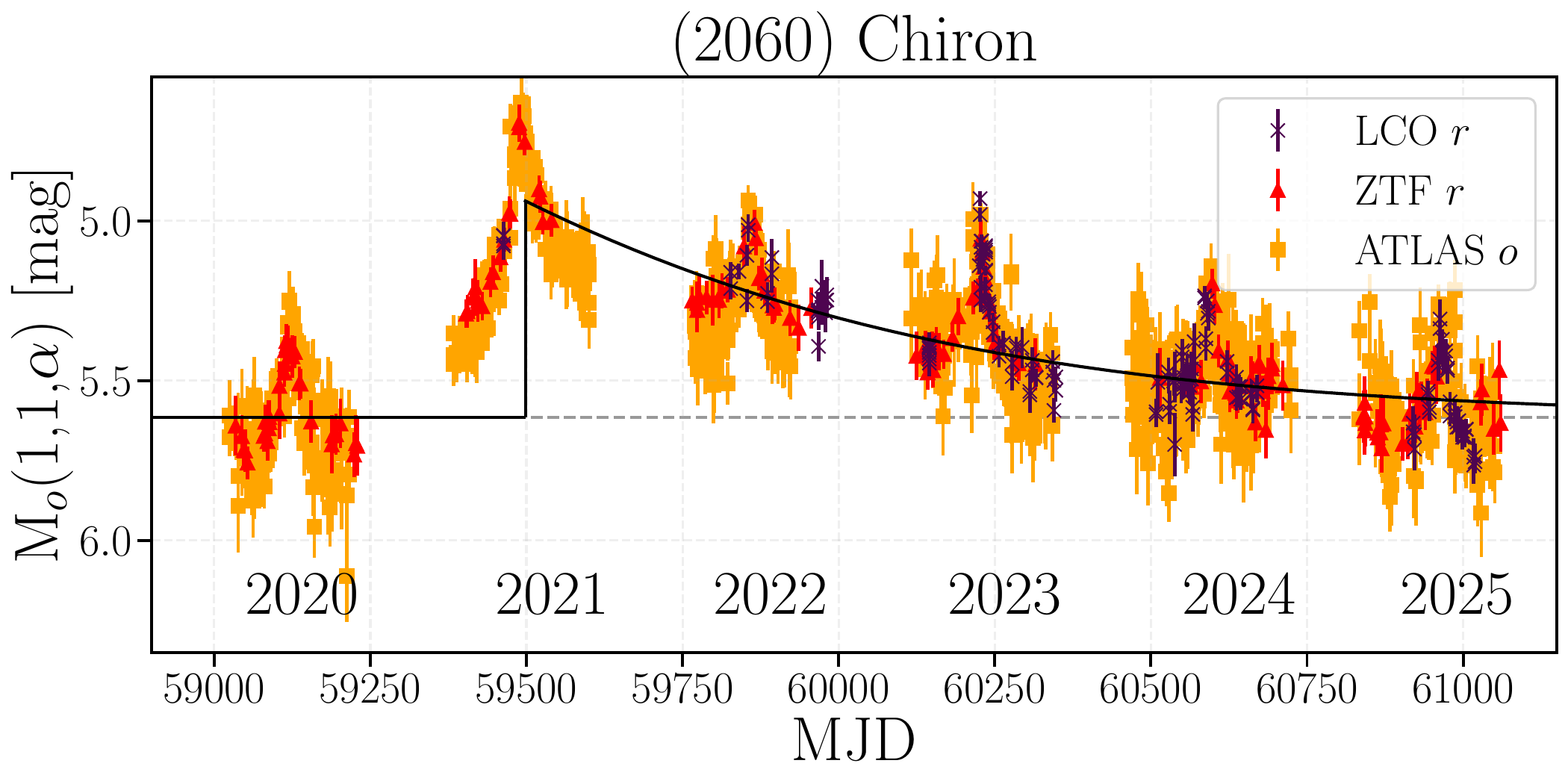}
    \caption{Distance-corrected light curves for Chiron color-shifted to ($top$) ATLAS $o$ band. The different datasets are denoted via differing colors. The solid black curve represents an exponential decay \update{fit to the 30-day binned median data} from 2021 post-opposition and onward. The gray dashed line represents the baseline quiescence magnitude, calculated as the median of 2020's observing season.}
    \label{fig:chiron_expdecay}
\end{figure*}

The long-term brightness evolution of Chiron, color-shifted to the ATLAS $c$ and $o$ filters, is shown in Figure \ref{chap4:fig:chiron_lc}, spanning 2020 to 2025. The sharp maxima in each season correspond to minimum phase angle observations. No explicit phase correction has been applied here, since Chiron's phase behavior has been seen to vary with time as a function of its activity evolution from 2021-2023 \citep{dobson24}. With two additional observing seasons in 2024 and 2025 beyond the dataset presented by \cite{dobson24}, it is now clear that the overall light curve morphology resembles a characteristic cometary outburst profile; a sharp brightening in 2021, followed by a steady decline towards the 2020 baseline. By 2024, the average reduced magnitude over the season was $M_o(1,1,\alpha)=5.51 \pm 0.07$ mag, consistent within $1\sigma$ with the average reduced magnitude over the 2020 season, $M_o(1,1,\alpha)=5.61\pm0.07$ mag, indicating that Chiron's brightness had largely returned to its pre-brightening level. To quantify the timescale of this relaxation, we model the post-opposition evolution in the color-shifted ATLAS-$o$ band data as a quiescent baseline of the median 2020 magnitude plus an exponentially decaying excess component, following a similar treatment in \cite{kelley21b}. \update{The data are binned into 30-day intervals and fitted simultaneously using least-squares minimization, yielding a decay timescale of $\tau = 1.38 \pm 0.16$~yr as shown in Figure~\ref{fig:chiron_expdecay}.} On this timescale, the excess brightness is expected to decay to 5\% of its initial amplitude after approximately \update{4.0~yr}, consistent with the observed convergence of the 2024-2026 photometry toward the 2020 baseline level.

\subsubsection{Mass-loss Estimates}
To place these photometric changes in a physical context, we convert the observed peak brightness in each observing season into an approximate upper mass-loss estimate, assuming ongoing dust production. We adopt a grain albedo of $p=0.18$ (c.f. ring modeling from \citealt{ortiz15, ortiz23, dobson24}), effective grain radius $a=1$ $\mu\text{m}$, density $\rho=0.5$~g~cm$^{-3}$ \citep[both per][]{fornasier13}, and velocity $v=100$~m~s$^{-1}$ \citep[as per estimates from 29P/Schwassmann-Wachmann 1,][]{wierzchos20}. The inferred dust mass-loss rates reach a maximum of $\lesssim81$~kg~s$^{-1}$ in 2021 pre-opposition, declining to $\lesssim53$~kg~s$^{-1}$ in 2023, and further to $\sim41$~kg~s$^{-1}$ by 2025. This is on the same order as the quiescent 2020 estimate of $\sim 43$~kg~s$^{-1}$, and only slightly lower than the $\sim50-100$~kg~s$^{-1}$ typically associated with weakly active Centaurs, such as 29P/Schwassmann-Wachmann 1 \citep{trigo08, trigo10, miles16}, demonstrating that the fading trend of Chiron is physically consistent with the decay of activity. 

An order-of-magnitude estimate of the dust mass associated with this excess brightness may be obtained by relating scattering cross-section to grain mass. Using the same grain parameters as above, the mass-to-cross-section ratio $\sfrac{M}{C}=\frac{4}{3}a\rho \approx 6.7\times10^{-4}$~kg~m$^{-2}$. The peak of the 2021 brightening corresponds to an enhancement of $\sim-0.7$~mag, equivalent to a flux increase of $\sfrac{F_{\rm peak}}{F_{2020}} \approx 1.9$. Assuming similar geometric albedo for the coma particles and the 2020 baseline scattering components, this implies the coma scattering cross-section is of order of the 2020 baseline cross-section itself, i.e., $\sfrac{C_{\rm coma}}{C_{2020}} \approx 0.9$. Assuming that the 2020 baseline cross-section is entirely due to Chiron's nucleus, this corresponds to $C \approx \pi R^2 \approx 2.5\times10^{10}$~m$^2$ for $R\approx90$~km \citep{bragaribas23}, and so $C_{\rm coma} \approx 10^{10}$~m$^2$. This places the ejecta mass of micron-sized grains at $\sim10^7$~kg, or $\sim10^8-10^9$~kg for larger grains. These values situate the 2021 brightening event within the range of ejected dust masses inferred for large cometary outbursts \citep{ishiguro16}, albeit towards the upper envelope relative to typical JFC events. Further, whilst subject to assumptions about the dust grain size, density, and velocity, which are poorly measured for Chiron, the mirroring of the secular decline in mass-loss supports the interpretation of the 2021 brightening event as a large impulsive brightening that has since decayed.  

\subsubsection{Phase Curve Evolution}\label{sec:4.2.3}
The secular phase curves further reinforce this picture. We fit 10$^5$ synthetic linear phase curves using \texttt{scipy.optimize.curve\_fit} to each observing season, offsetting each magnitude by a randomly drawn number from a Gaussian distribution with width of the uncertainty of that particular measurement. The obtained values for the absolute magnitude $H$ and linear phase slope parameter $\beta$ from this are summarized in Table \ref{tab:chiron_phase} in the ATLAS $o$ band. Uncertainties are quoted as 2$\sigma$ intervals for comparison to \cite{dobson24}. The 2021 brightening event saw Chiron achieve a significant pre-opposition brightening to $H_o=4.71\pm0.03$ mag, with a steep slope of $\beta_o=0.223\pm0.020$~mag~deg$^{-1}$. This value is significantly elevated relative to the quiescent 2020 coefficient $\beta_o = 0.108\pm0.021$~mag~deg$^{-1}$, consistent with the steepening reported by \cite{dobson24} during the same 2021 outburst epoch ($\beta_{o,\mathrm{Dobson}} = 0.226\pm0.008$~mag~deg$^{-1}$). Dust particles in an optically thin coma generally produce steeper opposition surges than a regolith surface due to enhanced coherent backscattering \citep{hapke93}\update{; bare cometary nuclei observed during inactive intervals and low-albedo C-type asteroids typically exhibit $\beta \approx 0.035-0.07$~mag~deg$^{-1}$ \citep{belskaya00, kokotanekova17}, well below Chiron's quiescent 2020 value and less than a third of the 2021 outburst slope.} Any coma present around Chiron appeared unresolved in stacked Gemini imaging \citep{dobson24}, which is consistent with an optically thin dust coma regime. The elevated phase slope therefore suggests a dust-driven phase behavior during the outburst — although we note that standard cometary phase curve models may not be directly applicable to Centaur comae.

In subsequent years, the absolute magnitude dimmed progressively by an average of $\sim0.1$ mag yr$^{-1}$, whilst the phase slope flattened reaching values comparable to the quiescent 2020 coefficient by $\sim2023-2025$. The post-outburst slopes measured between 2023 and 2025 are also mutually consistent within $2\sigma$, indicating a stabilization of the phase behavior as Chiron faded in brightness. Although the 2022 phase slope is somewhat lower than those of adjacent seasons, this deviation is not significant at the 2$\sigma$ level relative to the 2023 season. Given the limited phase angle ranges probed ($\sim0-3^\circ$, the linear phase slopes will be inherently sensitive to small ($\sim0.01$~mag) photometric scatter or offsets between datasets, which can produce fluctuations in the observed slope without requiring a distinct physical change occurring on Chiron. 

\begin{deluxetable}{c c c c c c c}
    \tablecaption{Absolute magnitudes and linear phase slope parameters obtained for Chiron from 2020-2025, color corrected to ATLAS \textit{o} bands.  \label{tab:chiron_phase}}
    \tablewidth{\columnwidth}
    \tablehead{
        \multirow{2}{*}{Observing Season} & $H_o$   & $\beta_o$          & $N_{obs,o}$ \\
                                          & [mag]   & [mag deg$^{-1}$]   &
    }
    \startdata
    2020                      & 5.43 $\pm$ 0.04 & 0.108 $\pm$ 0.021 & 160  \\
    2021\tablenotemark{\dag}  & 4.71 $\pm$ 0.03 & 0.223 $\pm$ 0.020 & 91  \\
    2021\tablenotemark{\ddag} & 4.69 $\pm$ 0.03 & 0.157 $\pm$ 0.020 & 102  \\
    2022                      & 5.15 $\pm$ 0.03 & 0.059 $\pm$ 0.016 & 226 \\
    2023                      & 5.22 $\pm$ 0.03 & 0.091 $\pm$ 0.017 & 184  \\
    2024                      & 5.33 $\pm$ 0.04 & 0.085 $\pm$ 0.018 & 306 \\
    2025                      & 5.49 $\pm$ 0.04 & 0.065 $\pm$ 0.021  & 173
    \enddata
    \tablecomments{Uncertainties are quoted as 2$\sigma$ for comparison with \cite{dobson24}.}
    \tablenotetext{\dag}{Pre-opposition data}
    \tablenotetext{\ddag}{Post-opposition data}
\end{deluxetable}

\begin{figure}
    \centering
    \includegraphics[width=1.0\linewidth]{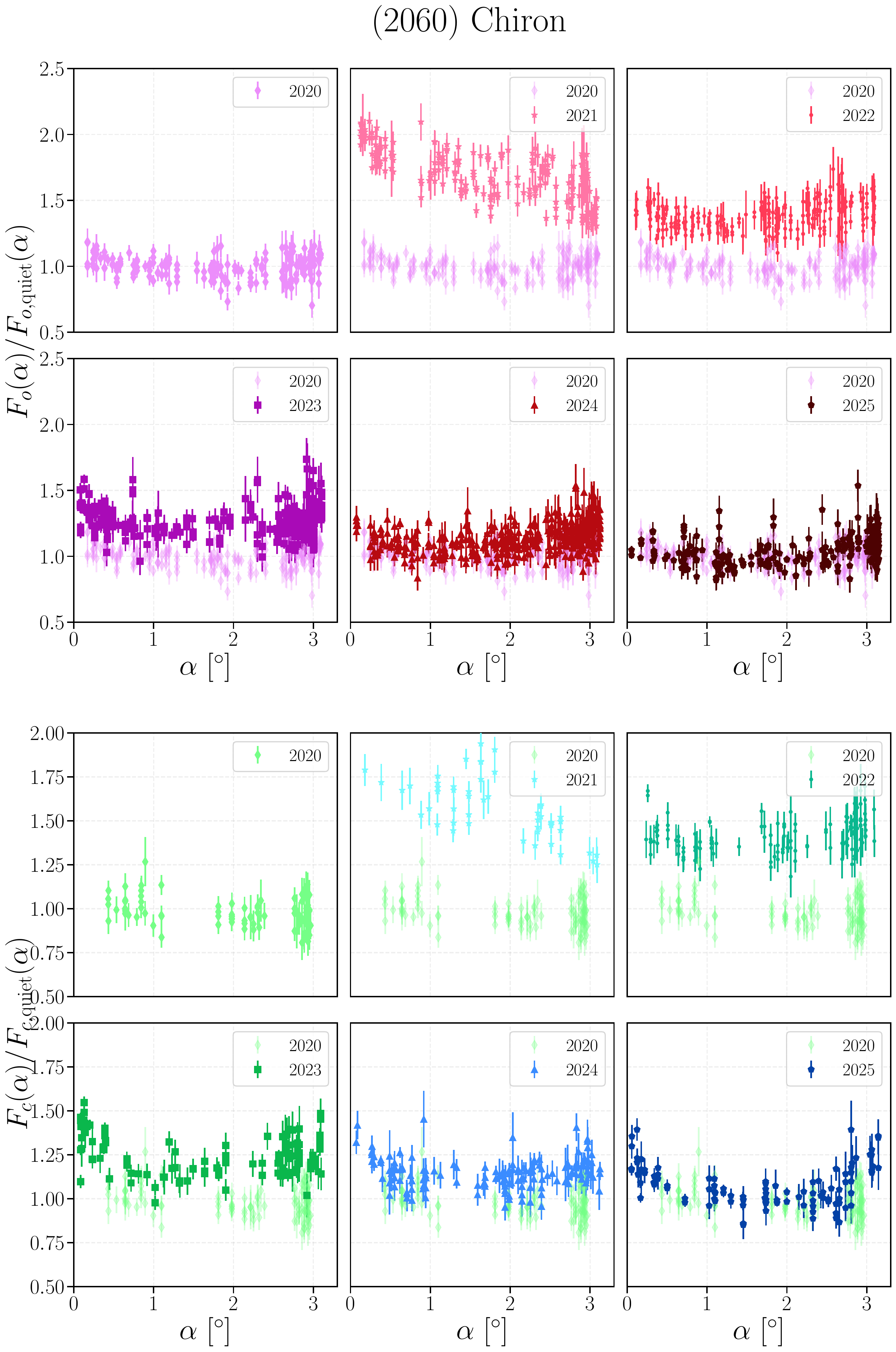}
    \caption{Residual phase curves, converted to flux space, for Chiron after subtracting off the quiescent 2015-2019 phase curve flux \citep{dobson24} for ($top$) the color-shifted to ATLAS $o$ band data, and ($bottom$) the color-shifted to ATLAS $c$ band data. Each panel represents a new observing season of data, with the 2020 phase curve included as a baseline for quiescent activity.}
    \label{chap4:fig:chiron_phase}
\end{figure}

To visualize how each season deviates from the quiescent behavior, Figure~\ref{chap4:fig:chiron_phase} shows the flux ratio $F(\alpha)/F_{\rm quiet}(\alpha)$ for both bands separately. For each individual data point within an observing season, the observed reduced magnitude M(1,1,$\alpha$) is first compared to the quiescent, pre-2020 linear phase curve, evaluated at that same phase angle. The resulting magnitude residual is converted to a linear flux ratio $F(\alpha)/F_{\rm quiet}(\alpha) = 10^{-0.4\left(M(\alpha) - M_{\rm quiet}(\alpha) \right)}$. In the quiescent 2020 observing season, the flux ratio will therefore remain approximately constant at unity. Deviations from unity in the later seasons quantify how the observed brightness differs from quiescent behavior at a given phase angle. If a given season differed only by an overall change in brightness (e.g.\ an increased effective scattering cross section), we would expect the flux ratio to remain approximately constant with phase angle. Instead, in Figure~\ref{chap4:fig:chiron_phase} both bands show a clear phase-dependence in their active years from 2021 to approximately 2023, with continued enhanced flux ratios at low phase angle that decline at larger phase angles. This would indicate that the evolution cannot be described by a simple scaling of the quiescent phase curve, but instead reflect changes in the phase dependence itself that progressively weaken in later years. Physically, this enhancement is consistent with additional scattering from coma particles (dust and/or icy grains). The progressive weakening of the phase dependence mirrors the secular fading of the absolute magnitude, and indicates that the additional particulate scattering present during 2021 had largely diminished by 2023-2025, approaching 2020-level quiescence. Notably, although the late-time brightness levels remain marginally elevated relative to 2020 in the 2024-2025 seasons (see Figure~\ref{chap4:fig:chiron_lc}), the phase coefficients from 2023-2025 are statistically indistinguishable at the $2\sigma$ level from that of 2020. This indicates that any residual excess flux contributes little additional phase dependence over the limited $\sim0-3^\circ$ phase angle range, perhaps consistent with a weakly scattering component superimposed on an otherwise quiescent nucleus/ring phase behavior.

Given that the Schleicher-Marcus phase function predicts only $\lesssim$0.1 mag of enhancement across the 0-3$^\circ$ phase range probed here, the steep 2021 (predominantly pre-opposition contribution) slope indicates scattering behavior stronger than expected for a typical cometary dust coma. The steep 2021 slope of $\beta_o=0.223\pm0.020$~mag~deg$^{-1}$, far exceeding both the Schleicher-Marcus prediction and Chiron's own quiescent baseline, is therefore most naturally attributed to enhanced backscattering from an optically thin coma of dust grains. Whilst the Schleicher-Marcus function provides a useful empirical reference, it was derived primarily interior to $\sim2$~au when cometary activity is typically driven by water ice sublimation and Halley dust observed at reduced phase angle sampling. The applicability of this prescription to Centaur comae, which may involve different grain properties, volatile drivers, dust compositions, or an intrinsically higher and evolving level of activity, therefore remains uncertain. 

\subsubsection{Physical Interpretation}
Cometary outbursts observed at heliocentric distances of a few au typically exhibit timescales of days to weeks \citep[e.g.][]{hughes90, fernandez05, ishiguro16, miles16b}. If the observed photometric evolution is dominated by material being transported out of a fixed angular aperture $\theta$, the characteristic timescale for this should scale as approximately $\tau_{\rm cross} \sim \frac{\theta \Delta}{v}$. Scaling typical cometary values ($\Delta \sim 1-2$~au, $v \sim 500$~m~s$^{-1}$) to Chiron's observing geometry ($\Delta \sim 18$~au, $v \sim 100$~m~s$^{-1}$) via $\sfrac{\tau_{\rm chiron}}{\tau_{\rm comet}} \sim \frac{\Delta_{\rm chiron}}{\Delta_{\rm comet}} \frac{v_{\rm comet}}{v_{\rm chiron}} \sim 90$. Under this scaling, a days-long cometary outburst maps to a months-year decay timescale, consistent with our measured $\tau \approx 1.40$~yr for Chiron. Additional effects likely contribute to this; Chiron's non-negligible escape velocity ($v_{\mathrm{esc}}\sim30-60$~m~s$^{-1}$, \citealt{dobson24}) is comparable to dust ejection velocities at larger heliocentric distances \citep[e.g.][]{wierzchos20}, meaning a fraction of ejecta may be gravitationally bound or only marginally unbound, leading to a persistent dust coma rather than freely dispersing ejecta. Additionally, whilst the ratio of radiation pressure to solar gravity on dust grains is constant with $r_\mathrm{H}$ \citep{burns79}, the absolute radiation pressure acceleration scales as ${r_\mathrm{H}}^{-2}$, making it $\sim340\times$ weaker at Chiron's $\sim18$~au distance than at 1~au, reducing its effectiveness at sweeping grains from Chiron's vicinity.

Beyond the geometric aperture argument, the physical conditions at Chiron's heliocentric distance further favor a gradual evolution. At Chiron's heliocentric distance ($r_\mathrm{H}\sim18.5$~au), water-ice sublimation is negligible, and activity is instead thought to be driven by alternative volatile species (CO, CO$_2$, CH$_4$; \citealt{jewitt09, womack17}). Compared to typical JFCs, gas production rates and dust velocities are generally observed to be lower at larger heliocentric distances \citep{wierzchos20, bockeleemorvan22}, and brightness evolution may therefore proceed more gradually. In this context, photometric variations evolving over months to years are more consistent with sustained or slowly declining activity, rather than the rapid fades characteristic of many near-Sun cometary outbursts. Consistent with this interpretation, JWST spectroscopy in \update{2023/24} revealed a volatile-rich coma emission and localized \update{CH$_4$ and CO$_2$ emission} \citep[the epoch of their observations is marked onto Figure~\ref{chap4:fig:chiron_lc} for reference,][]{pinillaalonso24, wong26}, demonstrating that whilst the continuum column density was low, \update{CO$_2$ or CH$_4$}-emission-driven activity was still present even as the phase curve had largely stabilized. Together, the fading of $H$ and the flattening of $\beta$ suggest that Chiron transitioned from a strongly scattering coma state in 2021 to a near-quiescent scattering regime by 2023-2025, despite evidence for continued low-level volatile activity.

However, the interpretation of Chiron's present brightness cannot rely on coma evolution alone, since its reported ring system also contributes to the observed flux \citep{bragaribas14, ortiz15, ortiz23, bragaribas23, sickafoose23}. \cite{dobson24} showed that Chiron has consistently been brighter than predictions from the two-ring model since at least 2014, with the 2021 brightening event exceeding the expected ring contribution by more than an order of magnitude, and thus requiring the explanation of a new epoch of activity. Our new epochs of photometry in this work have demonstrated that by 2024-2025, Chiron had approximately faded back to its 2020 brightness level - however this itself still lies $\sim0.5$ mag above the nominal ring-only prediction, which continues to decline with time. Combined with the JWST gas measurements, we argue that perhaps Chiron has not yet returned to a fully dormant state, but has instead entered a regime of low-level sustained activity, analogous with 29P/Schwassmann-Wachmann 1 between outbursts \citep{jewitt90, trigo08, miles16b, bockeleemorvan22, kareta25}. From this view, the 2021 brightening event therefore represents a large impulsive outburst event superimposed on a background of low-level coma production that has persisted into the present. Disentangling any potential residual comae from ring scattering is now crucial: only further high-cadence stellar occultations will directly constrain the evolving ring geometry and its azimuthal inhomogeneities \citep{ortiz23}, whilst repeat JWST imaging and spectroscopy can help to detect or rule out any further faint(er) dust comae. Together, these complementary approaches will be essential for establishing whether Chiron has truly settled into a new, long-term state of persistent weak activity, or if its variability can ultimately be reconciled within an updated (potentially three-ring-component, \citealt{bragaribas23, madeira25}) ring model framework.

\subsubsection{Color Evolution} \label{sec:4.2.2}
As with 103P, the near-simultaneous multi-filter coverage of Chiron within ATLAS, ZTF, and LCO allows us to obtain color measurements in order to probe any potential evolution in the coma color of Chiron. We employ the same methods used with 103P in Section \ref{sec:4.1.2} in order to measure ($c-o$) colors of Chiron, using both a $\pm2$~day nearest-neighbor pairing technique and a spline-based Monte Carlo approach implemented with \texttt{scipy.interpolate.UnivariateSpline} to fit $10^4$ synthetic light curves to the color-shifted ATLAS $c$ and $o$ band datasets within their measured magnitude uncertainties. Again, we only calculate corresponding magnitude measurements in each splined filter light curve where there is temporal overlap to ensure good quality spline fits to the data.

\begin{figure}
    \centering
    \includegraphics[width=0.95\linewidth]{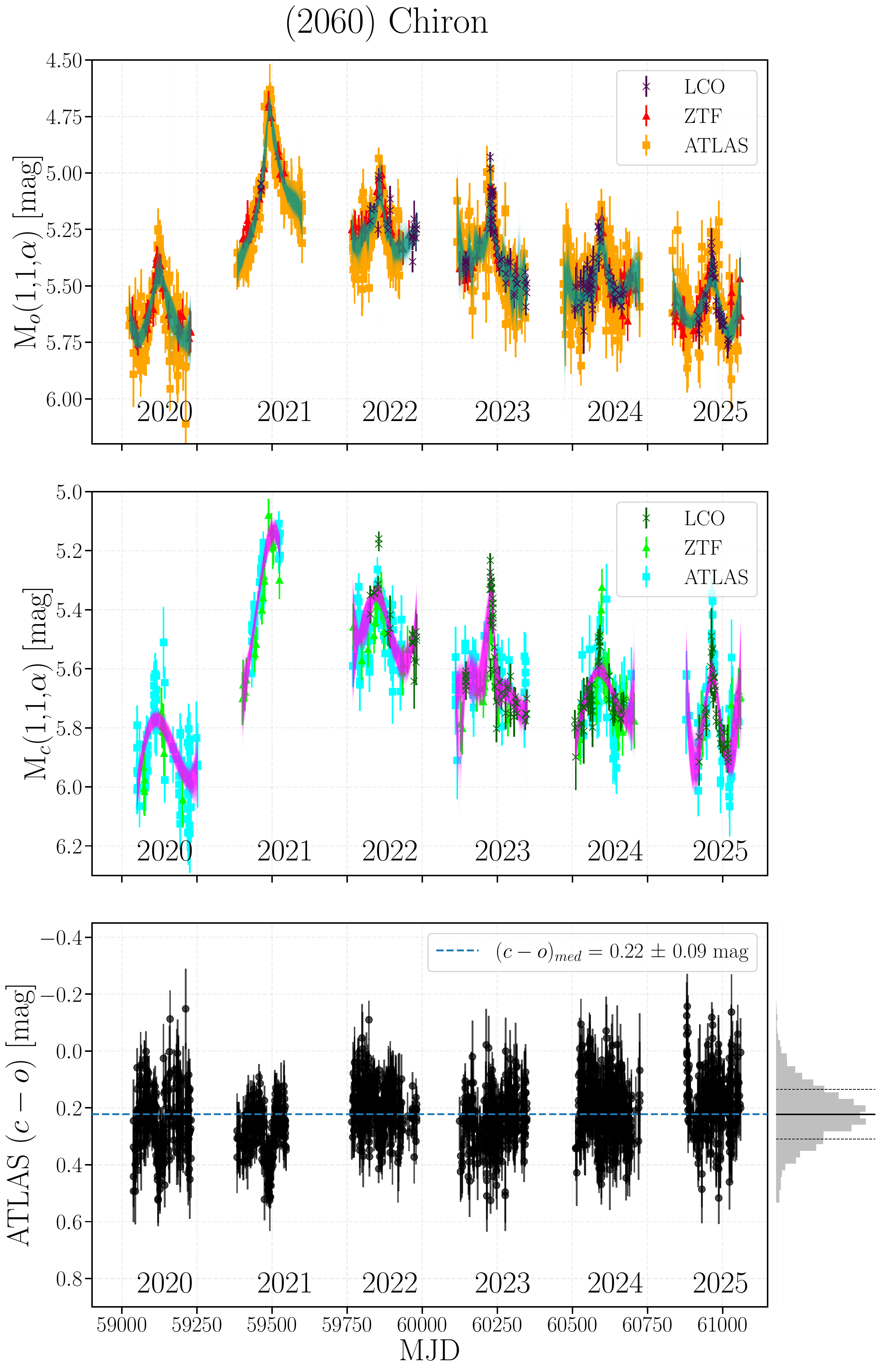}
    \caption{Light curves of the color-shifted magnitudes of Chiron to ($top$) the ATLAS $o$ band and ($middle$) the ATLAS $c$ band. Overplotted on each are $10^5$ synthetic Monte Carlo sampled splines within measurement uncertainties. ($bottom$) The color evolution of Chiron in ATLAS ($c-o$) over the shared time range, with uncertainty on each data point being estimated by the 1$\sigma$ uncertainties from the splines. The dashed horizontal line in blue represents the median colour across the dataset, and on the right is the histogram of the derived colors showing an approximate Gaussian distribution about the solid black line, with dashed black lines representing the standard deviation.} 
    \label{fig:chiron_colevo_spl}
\end{figure}

\begin{figure}
    \centering
    \includegraphics[width=0.9\linewidth]{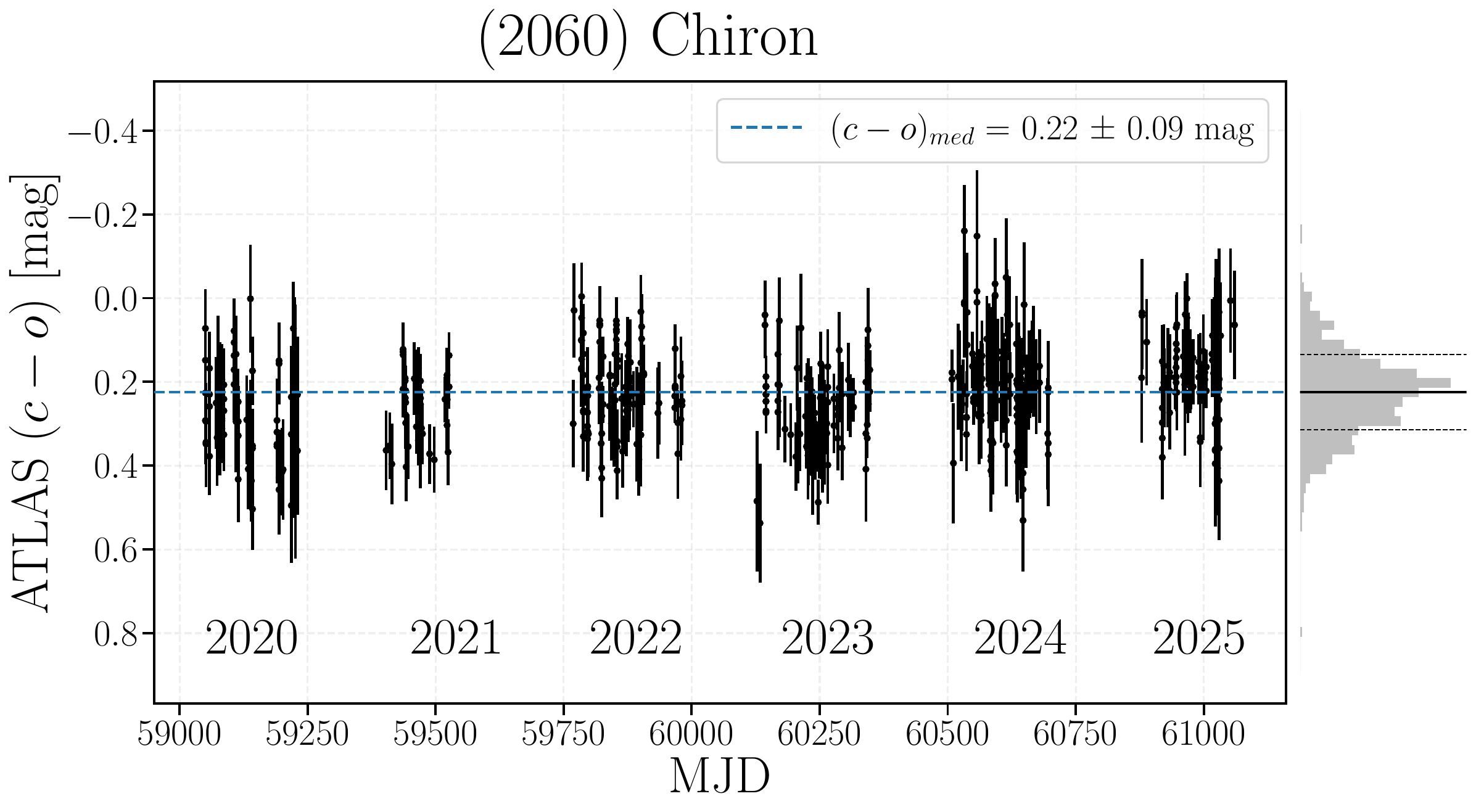}
    \caption{Color evolution of Chiron in ZTF ($g-r$) as in Figure~\ref{fig:chiron_colevo_spl}, this time using a nearest-neighbor approach with a time separation of $\pm$2 days. Uncertainties are obtained from propagating the individual photometric uncertainties in quadrature. The dashed horizontal line in blue, the histogram, and its associated annotations are the same as in Figure~\ref{fig:chiron_colevo_spl}.} 
    \label{fig:chiron_colevo_nn}
\end{figure}

The ($c-o$) color of Chiron derived from both the spline Monte Carlo method (Figure~\ref{fig:chiron_colevo_spl}) and the nearest-neighbor method (Figure~\ref{fig:chiron_colevo_nn}) shows no evidence for large-amplitude color variation across the entire 2020-2025 baseline, with ($c-o$) $=0.22\pm0.09$ mag. This trend is consistent with the 2017-2023 measurements from \cite{dobson24}, and with earlier epochs \citep{meech89, hartmann90, romonmartin03, fornasier04, fornasier09}, indicating that Chiron's broadband color has remained essentially unchanged before, during, and after its 2021 brightening event. The absence of any measurable color change suggests that the activity did not introduce a dust coma with differing optical properties from the nucleus, such as the cases of active Centaurs 174P/Echeclus \citep{bauer08, rousselot08, seccull19}, 29P/Schwassmann-Wachmann 1 \citep{jewitt09}, 166P/2001 T4 (NEAT) \citep{shi15}, and C/NEAT (2001 T4) \citep{bauer03b}. Instead, the additional flux was most likely driven by an increase in the amount of material surrounding Chiron. A small offset toward slightly bluer ($c-o$) values is visible in the 2021 block near opposition, corresponding to peak activity; however, the amplitude is modest relative to the overall seasonal scatter and is not obviously persistent across the season.

Such stability is however not unexpected for weakly active Centaurs; it has been discussed that the dust released during low-level activity can often be optically thin, allowing the nucleus to dominate the flux and preserve a stable color signature \citep[e.g.][]{luu90, jewitt09, epifani11, wong19}. However, the scale of Chiron's 2021 brightening and the deviation in linear phase slope parameter suggest that the coma's contribution was substantial, consistent with the $\sim0.7$~mag brightening observed in 2021, making it unlikely that the nucleus alone masked color changes. A more plausible explanation is that the coma dust shared broadly similar reflectance properties to the nucleus surface, such that even a bright coma did not alter the optical color. Earlier spectroscopy of a more quiescent Chiron found no measurable difference between coma and nucleus spectra \citep{meech89, romonmartin03, fornasier04, fornasier09}, whilst more recent JWST spectra from 2023 found no extended emission in a continuum slice between strong CH$_4$ emission bands \citep{pinillaalonso24}, potentially consistent with a dust component indistinguishable from the nucleus. Therefore, this continuation in color measurement strengthens the view of \cite{dobson24}, that the 2021 brightening event most likely reflects an increase in overall activity level, rather than the release of a compositionally distinct dust population. 

\update{\section{Future Prospects for the Legacy Survey of Space and Time (LSST)}\label{sec:5}}

\update{The NSF-DOE Vera C. Rubin Observatory's Legacy Survey of Space and Time \citep[LSST;][]{lsst09, schwamb18, ivezic19, bianco22} will represent a substantial advance over the survey capabilities demonstrated in this work. For 103P, the most concrete limitation of ATLAS and ZTF is one of depth: both surveys are unable to see 103P beyond $r_\mathrm{H} \sim$ 2.5-2.75~au as it approaches their limiting magnitudes. This leaves the large-heliocentric distance activity regime, which is known to persist to at least $\sim$5~au in prior apparitions \citep{snodgrass08}, entirely uncharacterized. The LSST's per-visit depth of $m_r \sim 24.7$~mag, approximately $100\times$ deeper in flux than ATLAS and $\sim40\times$ deeper than ZTF-$r$, will extend this coverage to larger heliocentric distances. For Chiron, which at $m \sim 18$~mag is comfortably within the reach of both current surveys, the limiting factor here is instead per-epoch photometric precision. Throughout this analysis, ATLAS's typical single-visit uncertainties of $\sim0.08-0.09$~mag and ZTF's $\sim0.07$~mag mean that phase curve slopes, activity levels, and outburst evolution can only be constrained through season-long averaging, washing out physical changes that may be occurring on shorter timescales. Single-visit photometric precision with the LSST at Chiron's brightness is expected to approach $\sim0.01$~mag \citep{ivezic19}, a factor of $\sim5-10\times$ improvement over the per-epoch uncertainties achieved by ATLAS and ZTF in this dataset.}

\update{Critically, the LSST applies these improvements across the entire accessible Centaur and JFC population simultaneously. Covering $\sim18000$~deg$^2$ of sky and revisiting each field every $\sim3$~nights \citep{bianco22}, a cadence broadly comparable to ATLAS and ZTF, over a nominal ten-year baseline, the LSST will monitor every detectable object continuously without requiring dedicated observing time. The multi-year characterization presented here for two objects will therefore become routine at population scale, and objects like Chiron can be interpreted in the context of a statistically meaningful sample \citep{murtagh25}.}

\update{The shared limitation that the LSST does not resolve is the inability to uniquely decompose gas emission from dust scattering within broadband photometry, which has limited interpretation throughout this work. For both 103P and Chiron, color analysis cannot uniquely separate molecular emission from dust scattering, or disentangle coma contributions from the nucleus. The LSST's wider, six-filter $ugrizy$ coverage will offer some additional leverage for population-level color diagnostics, but the fundamental ambiguity within any one filter persists regardless of photometric precision or depth. Dedicated narrowband photometry, stellar occultations, and spectroscopic follow-up will be required to confidently constrain gaseous contamination. The LSST is therefore best understood as being more than capable of providing the long-baseline, population-scale monitoring context into which such targeted campaigns can be embedded, rather than replacing them.}\\

\section{Conclusions} \label{sec:6}

In this study we have used wide-field survey datasets from ATLAS and ZTF, and targeted observations via LCO, to analyze the recent and long-term activity of 103P/Hartley~2 and (2060) Chiron through long-baseline light curves, phase curve analyses, broadband color measurements, and rotational variability searches.

For 103P/Hartley 2, the 2023/24 apparition exhibits pronounced perihelion asymmetry: steep inbound brightening ($n_r=-3.48\pm0.08$) and significantly flatter outbound fading ($n_r=-1.16\pm0.04$), consistent with the persistence of larger, slowly-moving dust grains whose lower radiation pressure susceptibility extends their coma contribution post-perihelion. Broadband color evolution shows a pronounced blueward excursion near perihelion, consistent with enhanced molecular emission contaminating the $g$-band, though heliocentric and phase-dependent effects preclude a unique gas/dust diagnosis from broadband data alone. A period search recovers a $\sim18.7$~hr activity-linked modulation near perihelion, consistent with \cite{knight15} and \cite{lehmann25}, but the survey cadence and the nucleus's complex NPA rotation state prevent an independent spin-down rate measurement or determination of the period's inter-apparition transition from $\sim19$~hr. \update{Together, these results place 103P at an erosive end-stage of JFC evolution. Its diminishing hyperactivity and increasingly dust-dominated post-perihelion coma are consistent with progressive volatile depletion across apparitions, in line with the secular fading reported by \cite{graykowski25}. Continued monitoring through its 2036/37 apparition will be needed to determine whether this decline continues monotonically, or is punctuated by episodic variability.}

For Chiron, two additional observing seasons capture the continued decay of the 2021 brightening toward the pre-brightening baseline, with phase slopes progressively returning to quiescent 2020 values by 2023-2025. The stability of the broadband ($c-o$) color throughout 2020-2025 implies that the evolving activity reflects changes in coma level rather than grain composition, consistent with the JWST detection of a volatile-rich gaseous coma with no continuum extension \citep{pinillaalonso24, wong26}. A persistent brightness offset above ring-only model predictions indicates Chiron may not have returned to a fully dormant state \citep[see][]{pereira25}, and continued stellar occultations and repeat JWST imaging will be essential for disentangling coma from ring contributions. \update{Chiron illustrates the complementary intermediate state in trans-Neptunian Object (TNO)-to-JFC evolution; still near aphelion, and retaining a volatile-rich inventory (CH$_4$, CO, CO$_2$), capable of sustaining episodic gas-dominated outbursts. The quiescent baseline established here is therefore critical for contextualizing Chiron's behavior as it approaches its 2046 perihelion, and for connecting the large-heliocentric distance activity regime to the kind of dust-dominated, volatile-depleted behavior exhibited by 103P.}

\update{The results for 103P and Chiron presented together in this work bracket the TNO-to-JFC evolutionary sequence. One object lies near the volatile-depleted end of that pathway, and the other is a volatile-rich Centaur still able to sustain episodic activity at large heliocentric distances. The long temporal baselines afforded by wide-field surveys are what make these evolutionary trends accessible in the first place, and will be crucial in contextualizing dedicated narrowband and spectroscopic campaigns. Continued coordinated monitoring of the objects in this study, especially with JWST's unique capacity to both detect the primary volatile drivers of cometary activity and characterize surface ices at large heliocentric distances \citep{kelley16}, remains the most direct path toward understanding small body evolution across this transition.}

\begin{acknowledgments}

J.M. acknowledges support from the Department for the Economy (DfE) Northern Ireland postgraduate studentship scheme. J.M. acknowledges support from the DiRAC Institute in the Department of Astronomy at the University of Washington. The DiRAC Institute is supported through generous gifts from the Charles and Lisa Simonyi Fund for Arts and Sciences, Janet and Lloyd Frink, and the Washington Research Foundation. J.M. thanks the LSST-DA Data Science Fellowship Program, which is funded by LSST-DA, the Brinson Foundation, and the Moore Foundation; his participation in the program has benefited this work. M.E.S. acknowledges support in part from UK Science and Technology Facilities Council (STFC) grant ST/X001253/1. M.M.D. was supported by the UK Science Technology Facilities Council (STFC) grant ST/V506990/1. R.R.L. was supported by the UK Science Technology Facilities Council (STFC) grant ST/V506990/1. A.F.G acknowledges support from the European Union's Horizon Europe Research and innovation programme under Grant Agreement No 101131928. Q.Y. is supported by NASA program 80NSSC21K0659.

The views expressed in this article are those of the authors and do not reflect the official policy or position of the U.S. Naval Academy, Department of the Navy, the Department of Defense, or the U.S. Government.

This work has made use of data from the Asteroid Terrestrial-impact Last Alert System (ATLAS) project. ATLAS is primarily funded to search for near-Earth asteroids through NASA grants NN12AR55G, 80NSSC18K0284, and 80NSSC18K1575; by-products of the NEO search include images and catalogs from the survey area. The ATLAS science products have been made possible through the contributions of the University of Hawaii Institute for Astronomy, the Queen's University Belfast, the Space Telescope Science Institute, the South African Astronomical Observatory (SAAO), and the Millennium Institute of Astrophysics (MAS), Chile.

This work is based on observations obtained with the Samuel Oschin Telescope 48-inch Telescope at the Palomar Observatory as part of the Zwicky Transient Facility project. ZTF is supported by the National Science Foundation under Grants No. AST-1440341, AST-2034437, and currently Award \#2407588. ZTF receives additional funding from the ZTF partnership. Current members include Caltech, USA; Caltech/IPAC, USA; University of Maryland, USA; University of California, Berkeley, USA; University of Wisconsin at Milwaukee, USA; Cornell University, USA; Drexel University, USA; University of North Carolina at Chapel Hill, USA; Institute of Science and Technology, Austria; National Central University, Taiwan, and OKC, University of Stockholm, Sweden. Operations are conducted by Caltech's Optical Observatory (COO), Caltech/IPAC, and the University of Washington at Seattle, USA.

This work makes use of observations from the Las Cumbres Observatory global telescope network. Observations with the LCOGT 1 m were obtained as part of the LCO Outbursting Objects Key (LOOK) project (KEY2020B-009).

This research has made use of services provided by NASA's Astrophysics Data System. This work has made use of data and services provided by the Horizons system of the Jet Propulsion Laboratory.

\update{We thank the anonymous reviewers for their comments on this work. Both of their insights helped to improved the quality of this manuscript.}

Data Access: All photometric data used in this study are provided in full as supplementary information accompanying this paper and are available in Appendix~\ref{ap:A}~and~\ref{ap:B}. Raw and Calibrated Observations from Las Cumbres Observatory used in this study are available at the LCO Science Archive (\href{https://archive.lco.global}{https://archive.lco.global}; proposal code KEY2020B-009) after an embargo/proprietary period of 12 months.

\end{acknowledgments}

\begin{contribution}

J.M. was responsible for writing and submitting the manuscript. He collated the archival datasets, scheduled LCO observations of Chiron for the 2023+2024+2025 blocks, and performed all of the analysis detailed in this manuscript. 

M.E.S supervised and contributed to the overall discussion and conception of the initial research concept. She provided meaningful edits and proofreads to this manuscript. 

M.S.P.K. reduced and provided ZTF photometry data for 103P/Hartley 2 used in this analysis, contributed to the overall discussion and conception of the research and provided meaningful edits and proofreads to this manuscript.

M.M.Knight, T.L., and G.M. provided high level discussion and feedback on the analysis of this work, and contributed to the conception of the initial research concept.

A.F.G. and A.F. took, reduced, and provided the ATLAS photometry for 103P/Hartley 2 used in this analysis.

M.M.D. scheduled the 2021+2022 LCO blocks of (2060) Chiron photometry for this work, whilst R.R.L. scheduled the latter half of the 2025 LCO block.

C.E.H., T.L., J.P.C., and S.G all contributed to building the NEOx platform which was used to schedule observations for the LCO dataset.

Q.Y., H.H.H., and D.B. all provided feedback on the text of the manuscript.

T.X.C., R.G.D., M.J.G., S.L.G., G.H., M.M. Kasliwal, A.A.M., and R.R. all contributed to the management, operations, and/or building of ZTF.


\end{contribution}

\facilities{ATLAS, PO:1.2m (ZTF), LCOGT (1-m telescopes)}

\software{
          Astropy \citep{astropy13, astropy18}, 
          Astroquery \citep{ginsburg19},
          JPL Horizons \citep{giorgini96},
          Jupyter Notebook \citep{kluyver16}
          Matplotlib \citep{hunter07},
          Numpy \citep{harris20}, 
          Pandas \citep{mckinney10, pandas20}, 
          sbpy \citep{mommert19},
          Scipy \citep{virtanen20},
          seaborn \citep{waskom21},
          tqdm \citep{dacostaluis23}
          }

\appendix
\restartappendixnumbering

\section{Observations of 103P/Hartley 2} \label{ap:A}
The observations of 103P/Hartley 2 used in this work are broken down by survey/telescope and listed in Tables~\ref{ap:A:tab1}-\ref{ap:A:tab3}

\begin{deluxetable*}{cccccccccccc}
    \tablecaption{ATLAS observations of 103P \label{ap:A:tab1}}
    \tablehead{
        \multirow{2}{*}{UT Date} & \multirow{2}{*}{MJD} & \colhead{Magnitude} & \colhead{Distance Corrected Magnitude} & \colhead{Absolute Magnitude} & \colhead{Magnitude Error} & \multirow{2}{*}{Filter} & \multirow{2}{*}{$r_\mathrm{H}$\tablenotemark{a}} & \multirow{2}{*}{$\Delta$\tablenotemark{b}} & \multirow{2}{*}{$\alpha$\tablenotemark{c}} & \multirow{2}{*}{Site\tablenotemark{d}} \\ 
        & & \colhead{$m_\lambda(r_\mathrm{H},\Delta,\alpha)$} & \colhead{$m_\lambda(1,1,\alpha)$} & \colhead{$m_\lambda(1,1,0)$} & \colhead{$\sigma_m$} & & & & &\\
        & & & & & & & \colhead{(au)} & \colhead{(au)} & \colhead{(deg)} &
    } 
    \startdata
    2023 May 28.01 & 60092.008519 & 18.580 & 15.930 & 14.967 & 0.173 & c & 2.005 & 1.690 & 30.334 & SAAO \\
    2023 May 28.01 & 60092.011344 & 18.498 & 15.848 & 14.885 & 0.159 & c & 2.005 & 1.690 & 30.335 & SAAO \\
    2023 May 28.02 & 60092.016557 & 18.632 & 15.982 & 15.019 & 0.175 & c & 2.005 & 1.690 & 30.336 & SAAO \\
    2023 May 28.05 & 60092.052350 & 18.460 & 15.811 & 14.848 & 0.134 & c & 2.005 & 1.689 & 30.341 & SAAO \\
    2023 Jun 6.06  & 60101.055239 & 18.337 & 15.993 & 15.007 & 0.307 & o & 1.925 & 1.529 & 31.600 & SAAO \\
    2023 Jun 13.34 & 60108.339690 & 18.069 & 15.985 & 14.982 & 0.165 & o & 1.860 & 1.404 & 32.605 & CHL \\
    2023 Jun 13.34 & 60108.342433 & 17.726 & 15.642 & 14.639 & 0.124 & o & 1.860 & 1.404 & 32.606 & CHL \\
    \enddata
    \tablecomments{All observations are taken with 30~s exposure time}
    \tablenotetext{a}{Heliocentric distance}
    \tablenotetext{b}{Geocentric distance}
    \tablenotetext{c}{Phase angle}
    \tablenotetext{d}{HKO (Haleakal\={a} Hawai'i), MLO (Mauna Loa, Hawai'i), SAAO (South African Astronomical Observatory, South Africa), CHL (R\'io Hurtado, Chile), TDO (Teide Observatory, Tenerife)}
    {This table of observations is published in machine-readable format in its entirety in the online article. A portion is shown here as an example of its form and content.}
\end{deluxetable*}
 
\begin{deluxetable*}{ccccccccccc}
    \tablecaption{ZTF observations of 103P \label{ap:A:tab2}}
    \tablehead{
        \multirow{2}{*}{UT Date} & \multirow{2}{*}{MJD} & \colhead{Magnitude} & \colhead{Distance Corrected Magnitude} & \colhead{Absolute Magnitude} & \colhead{Magnitude Error} & \multirow{2}{*}{Filter} & \multirow{2}{*}{$r_\mathrm{H}$\tablenotemark{a}} & \multirow{2}{*}{$\Delta$\tablenotemark{b}} & \multirow{2}{*}{$\alpha$\tablenotemark{c}}  \\ 
        & & \colhead{$m_\lambda(r_\mathrm{H},\Delta,\alpha)$} & \colhead{$m_\lambda(1,1,\alpha)$} & \colhead{$m_\lambda(1,1,0)$} & \colhead{$\sigma_m$} & & & & \\
        & & & & & & & \colhead{(au)} & \colhead{(au)} & \colhead{(deg)}
    } 
    \startdata
        2023 Jun 6.41   & 60101.410648 & 18.360 & 16.028 & 15.042 & 0.207 & r & 1.922 & 1.523 & 31.648 \\
        2023 Jun 6.41   & 60101.411123 & 18.412 & 16.080 & 15.093 & 0.211 & r & 1.922 & 1.523 & 31.648 \\
        2023 Jun 8.41   & 60103.406782 & 18.184 & 15.922 & 14.930 & 0.153 & r & 1.904 & 1.488 & 31.923 \\
        2023 Jun 8.41   & 60103.409178 & 18.155 & 15.893 & 14.902 & 0.164 & r & 1.904 & 1.488 & 31.924 \\
        2023 Jun 8.43   & 60103.434120 & 18.144 & 15.883 & 14.892 & 0.242 & g & 1.904 & 1.488 & 31.927 \\
        2023 Jun 15.42  & 60110.415104 & 18.285 & 16.275 & 15.268 & 0.067 & g & 1.842 & 1.370 & 32.889 \\
        2023 Jun 15.45  & 60110.445185 & 17.905 & 15.896 & 14.889 & 0.049 & r & 1.842 & 1.369 & 32.893 \\
    \enddata
    \tablecomments{All observations are taken with 30~s exposure time}
    \tablenotetext{a}{Heliocentric distance}
    \tablenotetext{b}{Geocentric distance}
    \tablenotetext{c}{Phase angle}
    {This table of observations is published in machine-readable format in its entirety in the online article. A portion is shown here as an example of its form and content}
\end{deluxetable*}
 
\begin{deluxetable*}{ccccccccccccccc}
    \tablecaption{LCO observations of 103P \label{ap:A:tab3}}
    \tablehead{
        \multirow{2}{*}{UT Date} & \multirow{2}{*}{MJD} & \colhead{Magnitude} & \colhead{Distance Corrected Magnitude} & \colhead{Absolute Magnitude} & \colhead{Magnitude Error} & \multirow{2}{*}{Filter} & \multirow{2}{*}{$r_\mathrm{H}$\tablenotemark{a}} & \multirow{2}{*}{$\Delta$\tablenotemark{b}} & \multirow{2}{*}{$\alpha$\tablenotemark{c}} & \multirow{2}{*}{$t_{exp}$\tablenotemark{d}} & \multirow{2}{*}{Seeing} & \multirow{2}{*}{Airmass}  \\ 
        & & \colhead{$m_\lambda(r_\mathrm{H},\Delta,\alpha)$} & \colhead{$m_\lambda(1,1,\alpha)$} & \colhead{$m_\lambda(1,1,0)$} & \colhead{$\sigma_m$} & & & & & & & \\
        & & & & & & & \colhead{(au)} & \colhead{(au)} & \colhead{(deg)} & \colhead{(s)} & \colhead{(arcsec)} &
    }
    \startdata
        2023 Sep 24.37 & 60211.372222 & 12.163 & 14.054 & 12.952 & 0.023 & r & 1.093 & 0.383 & 66.399 & 82 & 2.93 & 1.415 \\
        2023 Sep 24.37 & 60211.372917 & 11.824 & 13.715 & 12.613 & 0.019 & g & 1.093 & 0.383 & 66.399 & 82 & 2.76 & 1.413 \\
        2023 Sep 26.11 & 60213.111111 & 11.995 & 13.896 & 12.801 & 0.044 & r & 1.088 & 0.383 & 67.059 & 83 & 2.50 & 1.637 \\
        2023 Sep 26.11 & 60213.111806 & 11.881 & 13.782 & 12.687 & 0.037 & g & 1.088 & 0.383 & 67.058 & 83 & 2.85 & 1.635 \\
        2023 Sep 27.17 & 60214.168750 & 12.013 & 13.920 & 12.829 & 0.030 & r & 1.085 & 0.383 & 67.426 & 83 & 3.89 & 1.239 \\
        2023 Sep 27.17 & 60214.168750 & 11.717 & 13.624 & 12.533 & 0.028 & g & 1.085 & 0.383 & 67.426 & 83 & 4.09 & 1.238 \\
        2023 Sep 28.12 & 60215.122917 & 12.070 & 13.981 & 12.893 & 0.027 & r & 1.083 & 0.383 & 67.738 & 83 & 2.56 & 1.571 \\
    \enddata
    \tablenotetext{a}{Heliocentric distance}
    \tablenotetext{b}{Geocentric distance}
    \tablenotetext{c}{Phase angle}
    \tablenotetext{d}{Exposure time}
    {This table of observations is published in machine-readable format in its entirety in the online article. A portion is shown here as an example of its form and content}
\end{deluxetable*}

\section{Observations of (2060) Chiron} \label{ap:B}
The observations of (2060) Chiron used in this work are broken down by survey/telescope and listed in Tables~\ref{ap:B:tab1}-\ref{ap:B:tab3}

\begin{deluxetable*}{ccccccccccc}
    \tablecaption{ATLAS observations of Chiron \label{ap:B:tab1}}
    \tablehead{
        \multirow{2}{*}{UT Date} & \multirow{2}{*}{MJD} & \colhead{Magnitude} & \colhead{Distance Corrected Magnitude} & \colhead{Absolute Magnitude} & \colhead{Magnitude Error} & \multirow{2}{*}{Filter} & \multirow{2}{*}{$r_\mathrm{H}$\tablenotemark{a}} & \multirow{2}{*}{$\Delta$\tablenotemark{b}} & \multirow{2}{*}{$\alpha$\tablenotemark{c}} & \multirow{2}{*}{Site\tablenotemark{d}} \\ 
        & & \colhead{$m_\lambda(r_\mathrm{H},\Delta,\alpha)$} & \colhead{$m_\lambda(1,1,\alpha)$} & \colhead{$m_\lambda(1,1,0)$} & \colhead{$\sigma_m$} & & & & &\\
        & & & & & & & \colhead{(au)} & \colhead{(au)} & \colhead{(deg)} &
    } 
    \startdata
    2020 Jun 24.60 & 59024.601392 & 18.438 & 5.676 & 5.541 & 0.060 & o & 18.851 & 18.926 & 3.076 & HKO \\
    2020 Jun 28.57 & 59028.565901 & 18.395 & 5.641 & 5.505 & 0.042 & o & 18.851 & 18.860 & 3.090 & HKO \\
    2020 Jun 28.57 & 59028.571430 & 18.404 & 5.650 & 5.514 & 0.059 & o & 18.851 & 18.860 & 3.090 & HKO \\
    2020 Jun 28.59 & 59028.591528 & 18.398 & 5.644 & 5.508 & 0.065 & o & 18.851 & 18.860 & 3.090 & HKO \\
    2020 Jul 2.60 & 59032.604220 & 18.389 & 5.642 & 5.507 & 0.059 & o & 18.852 & 18.793 & 3.091 & HKO \\
    2020 Jul 2.61 & 59032.614355 & 18.400 & 5.653 & 5.518 & 0.071 & o & 18.852 & 18.793 & 3.091 & HKO \\
    2020 Jul 3.58 & 59033.581424 & 18.348 & 5.603 & 5.468 & 0.086 & o & 18.852 & 18.777 & 3.090 & HKO \\
    \enddata
    \tablecomments{All observations are taken with 30~s exposure time}
    \tablenotetext{a}{Heliocentric distance}
    \tablenotetext{b}{Geocentric distance}
    \tablenotetext{c}{Phase angle}
    \tablenotetext{d}{HKO (Haleakal\={a} Hawai'i), MLO (Mauna Loa, Hawai'i), SAAO (South African Astronomical Observatory, South Africa), CHL (R\'io Hurtado, Chile), TDO (Teide Observatory, Tenerife)}
    {This table of observations is published in machine-readable format in its entirety in the online article. A portion is shown here as an example of its form and content.}
\end{deluxetable*}

\begin{deluxetable*}{cccccccccc}
    \tablecaption{ZTF observations of Chiron \label{ap:B:tab2}}
    \tablehead{
        \multirow{2}{*}{UT Date} & \multirow{2}{*}{MJD} & \colhead{Magnitude} & \colhead{Distance Corrected Magnitude} & \colhead{Absolute Magnitude} & \colhead{Magnitude Error} & \multirow{2}{*}{Filter} & \multirow{2}{*}{$r_\mathrm{H}$\tablenotemark{a}} & \multirow{2}{*}{$\Delta$\tablenotemark{b}} & \multirow{2}{*}{$\alpha$\tablenotemark{c}}  \\ 
        & & \colhead{$m_\lambda(r_\mathrm{H},\Delta,\alpha)$} & \colhead{$m_\lambda(1,1,\alpha)$} & \colhead{$m_\lambda(1,1,0)$} & \colhead{$\sigma_m$} & & & & \\
        & & & & & & & \colhead{(au)} & \colhead{(au)} & \colhead{(deg)}
    } 
    \startdata
        2020 Jul 4.38    & 59034.382025 & 18.407 & 5.664 & 5.528 & 0.094 & r & 18.852 & 18.764 & 3.086 \\
        2020 Jul 14.46   & 59044.460741 & 18.420 & 5.696 & 5.564 & 0.065 & r & 18.853 & 18.598 & 3.011 \\
        2020 Jul 17.42   & 59047.419954 & 18.458 & 5.739 & 5.609 & 0.050 & r & 18.853 & 18.550 & 2.973 \\
        2020 Jul 19.39   & 59049.393426 & 18.456 & 5.741 & 5.612 & 0.065 & r & 18.854 & 18.519 & 2.943 \\
        2020 Jul 23.46   & 59053.459907 & 18.487 & 5.768 & 5.640 & 0.051 & r & 18.854 & 18.455 & 2.871 \\
        2020 Aug 11.48 & 59072.478079 & 18.367 & 5.691 & 5.587 & 0.059 & r & 18.856 & 18.186 & 2.354 \\
        2020 Aug 12.41 & 59073.410058 & 18.819 & 6.116 & 5.990 & 0.087 & g & 18.856 & 18.174 & 2.322 \\
    \enddata
    \tablecomments{All observations are taken with 30~s exposure time}
    \tablenotetext{a}{Heliocentric distance}
    \tablenotetext{b}{Geocentric distance}
    \tablenotetext{c}{Phase angle}
    {This table of observations is published in machine-readable format in its entirety in the online article. A portion is shown here as an example of its form and content}
\end{deluxetable*}

\begin{deluxetable*}{ccccccccccccc}
    \tablecaption{LCO observations of Chiron \label{ap:B:tab3}}
    \tablehead{
        \multirow{2}{*}{UT Date} & \multirow{2}{*}{MJD} & \colhead{Magnitude} & \colhead{Distance Corrected Magnitude} & \colhead{Absolute Magnitude} & \colhead{Magnitude Error} & \multirow{2}{*}{Filter} & \multirow{2}{*}{$r_\mathrm{H}$\tablenotemark{a}} & \multirow{2}{*}{$\Delta$\tablenotemark{b}} & \multirow{2}{*}{$\alpha$\tablenotemark{c}} & \multirow{2}{*}{$t_{exp}$\tablenotemark{d}} & \multirow{2}{*}{Seeing} & \multirow{2}{*}{Airmass}  \\ 
        & & \colhead{$m_\lambda(r_\mathrm{H},\Delta,\alpha)$} & \colhead{$m_\lambda(1,1,\alpha)$} & \colhead{$m_\lambda(1,1,0)$} & \colhead{$\sigma_m$} & & & & & & & \\
        & & & & & & & \colhead{(au)} & \colhead{(au)} & \colhead{(deg)} & \colhead{(s)} & \colhead{(arcsec)} &
    }
    \startdata
        2021 Sep 6.32    & 59463.321285 & 17.802 & 5.151 & 5.086 & 0.044 & r & 18.868 & 17.971 & 1.442 & 245 & 1.73 & 1.344 \\
        2021 Sep 6.32   & 59463.324421 & 17.795 & 5.144 & 5.079 & 0.051 & r & 18.868 & 17.971 & 1.442 & 245 & 1.68 & 1.355 \\
        2021 Sep 6.33   & 59463.327569 & 17.771 & 5.120 & 5.055 & 0.042 & r & 18.868 & 17.971 & 1.442 & 245 & 1.62 & 1.367 \\
        2021 Sep 6.33   & 59463.330706 & 17.769 & 5.118 & 5.053 & 0.043 & r & 18.868 & 17.971 & 1.441 & 245 & 1.61 & 1.380 \\
        2022 Sep 5.02   & 59827.018935 & 18.178 & 5.529 & 5.455 & 0.033 & g & 18.834 & 17.981 & 1.682 & 180 & 1.92 & 1.347 \\
        2022 Sep 5.02 & 59827.021377 & 17.936 & 5.287 & 5.213 & 0.034 & r & 18.834 & 17.981 & 1.682 & 180 & 2.06 & 1.332 \\
        2022 Sep 5.02 & 59827.023831 & 18.241 & 5.592 & 5.518 & 0.032 & g & 18.834 & 17.981 & 1.682 & 180 & 2.01 & 1.318 \\
    \enddata
    \tablenotetext{a}{Heliocentric distance}
    \tablenotetext{b}{Geocentric distance}
    \tablenotetext{c}{Phase angle}
    \tablenotetext{d}{Exposure time}
    {This table of observations is published in machine-readable format in its entirety in the online article. A portion is shown here as an example of its form and content}
\end{deluxetable*}

\bibliography{zrefs}{}
\bibliographystyle{aasjournalv7}

\end{CJK*}
\end{document}